\documentclass[11pt]{article}
\usepackage[verbose=true,letterpaper]{geometry}
\AtBeginDocument{
  \newgeometry{
    textheight=9.2in,
    textwidth=6.9in,
    top=0.9in,
    headheight=14pt,
    headsep=25pt,
    footskip=30pt
  }
}
\usepackage{footnote}

\usepackage{natbib} 
\usepackage[utf8]{inputenc}
\usepackage{setspace}
\usepackage{amsmath,epsfig,epsf,psfrag, url, setspace, graphicx,natbib,mathtools,subfigure,comment}
\usepackage{a4wide,amsmath, amsthm, amssymb, multirow}
\usepackage{xcolor,colortbl}
\usepackage{amsfonts}
\usepackage{verbatim}
\usepackage{arydshln} 
\usepackage{placeins} 
\providecommand{\keywords}[1]
{
  \small	
  \textbf{\textit{Keywords---}} #1
}
\usepackage{bm}
\usepackage{makecell}
\usepackage[flushleft]{threeparttable} 
\usepackage{bm}
\usepackage{graphicx}
\usepackage{placeins}
\usepackage{amsmath}
\usepackage{amsfonts} 
\usepackage{commath} 
\usepackage{stackengine}
\usepackage[T1]{fontenc} 
\stackMath
\newcommand\tsup[2][2]{
 \def\useanchorwidth{T}%
  \ifnum#1>1%
    \stackon[-.5pt]{\tsup[\numexpr#1-1\relax]{#2}}{\scriptscriptstyle\sim}%
  \else%
    \stackon[.5pt]{#2}{\scriptscriptstyle\sim}%
  \fi%
}
\usepackage{tabularx, booktabs}
\usepackage{multirow}
\usepackage[flushleft]{threeparttable}
\usepackage{adjustbox}
\usepackage{appendix}

\usepackage[linesnumbered,ruled,vlined]{algorithm2e}
\usepackage[noend]{algpseudocode}
\newcolumntype{Y}{>{\centering\arraybackslash}X}

\title{Bayesian Methods for the Analysis of Early-Phase Oncology Basket Trials with Information Borrowing across Cancer Types}

\author{Jin Jin$^{1}$\footnotemark[1], Marie-Karelle Riviere$^2$, Xiaodong Luo$^3$,  Yingwen Dong$^4$\\\\
\vspace{6pt}
$^1$Department of Biostatistics, Johns Hopkins Bloomberg\\School of Public Health, Baltimore, MD, USA\\\\
\vspace{2pt}
$^2$Department of Biostatistics and Programming, Research and Development,\\Sanofi, Chilly-Mazarin, France.\\\\
\vspace{2pt}
$^3$Department of Biostatistics and Programming,\\Sanofi, Bridgewater, NJ, U.S.\\\\
\vspace{2pt}
$^4$Department of Biostatistics and Programming Oncology,\\Sanofi,  Cambridge, MA, U.S.}
\date{}

\begin{document}
\maketitle

\footnotetext{$^{\ast}$To whom correspondence should be addressed. E-mail: jjin31@jhu.edu}


\begin{abstract}
Research in oncology has changed the focus from histological properties of tumors in a specific organ to a specific genomic aberration potentially shared by multiple cancer types. This motivates the basket trial, which assesses the efficacy of treatment simultaneously on multiple cancer types that have a common aberration. Although the assumption of homogeneous treatment effects seems reasonable given the shared aberration, in reality, the treatment effect may vary by cancer type, and potentially only a subgroup of the cancer types respond to the treatment. Various approaches have been proposed to increase the trial power by borrowing information across cancer types, which, however, tend to inflate the type I error rate. In this paper, we review some representative Bayesian information borrowing methods for the analysis of early-phase basket trials. We then propose a novel method called the Bayesian hierarchical model with a correlated prior (CBHM), which conducts more flexible borrowing across cancer types according to sample similarity. We did simulation studies to compare CBHM with independent analysis and three information borrowing approaches: the conventional Bayesian hierarchical model, the EXNEX approach and Liu's two-stage approach. Simulation results show that all information borrowing approaches substantially improve the power of independent analysis if a large proportion of the cancer types truly respond to the treatment. Our proposed CBHM approach shows an advantage over the existing information borrowing approaches, with a power similar to that of EXNEX or Liu's approach, but the potential to provide substantially better control of type I error rate.
\end{abstract}



\section{Introduction}\label{intro}
Traditional oncology therapeutic research and drug development have mainly focused on the histological properties of tumors in each specific organ. Now with enhanced technology in biomarker development and genomic medicine, the focus has shifted from conventional chemotherapy to targeted therapy on a specific genomic or molecular aberration, which can potentially be shared by multiple histological cancer types in different organs.  
Under this new paradigm, a recently emerged type of trial called basket trial has received increasing attention. A basket trial enrolls patients with multiple histological cancer types (indications) that share the same genomic or molecular aberration, and assesses the efficacy of a particular targeted therapy simultaneously on all of the enrolled indications.
Previous research has shown that basket trials tend to require fewer patients and shorter trial duration to identify indications that respond to the treatment; 
in addition, it has the advantage of allowing patients with certain rare cancer types to be eligible to participate in clinical trials (Redig and J{\"a}nne, 2015; Chu and Yuan, 2018).

Typically, in the early, exploratory phase of oncology basket trials, only a small number of patients are enrolled for each indication, thus assessing the treatment effect on each indication independently is inefficient with low statistical power. 
In a basket trial, all indications share the same genetic or genomic aberration, leading to a prior guess of homogeneous treatment effects. 
Based on this exchangeability (EX) assumption, the idea of borrowing information across indications was introduced by the use of Bayesian hierarchical model (BHM) to improve trial efficiency, where the treatment effects are assumed to be exchangeable and centering around a common mean (Thall et al., 2013; Berry et al., 2013). 
In reality, however, the EX assumption is often violated, and it is common that only a subset of the indications are sensitive (i.e. the patients respond positively) to the treatment. 
For example, the MyPathway study shows that the response rate to trastuzumab plus pertuzumab differs greatly between several HER2-amplified/overexpressing tumor types (Hainsworth et al., 2018). 
In this case, BHM tends to over-shrink treatment effects toward the common mean, which can lead to highly inflated type I error rate for the insensitive indications (Freidlin and Korn, 2013). 

Recently, various information borrowing approaches have been proposed for early-phase basket trials to improve the power of independent analysis and, at the same time, provide better type I error control than BHM. 
However, no formal investigation has been conducted on the relative performance of these existing approaches.
In this paper, we discuss two recently published approaches. 
One is the exchangeability-nonexchangeability (EXNEX) method (Neuenschwander et al., 2016),  
which is a robust extension of the conventional BHM that assumes that each indication is either exchangeable (EX) with some indications or nonexchangeable (NEX) with any other indication, i.e. an outlier. 
The other is Liu's two-path method, which introduces an interim homogeneity test, and only conducts information borrowing when there is not enough evidence to reject the homogeneity hypothesis (Liu et al., 2017). 
This homogeneity test is introduced for better type I error control. However, the doubt raises regarding the necessity of the test because it introduces additional tuning parameters and complicates the trial design. 
In addition, the evidence for heterogeneity does not necessarily indicate that all indications are heterogeneous and should be analyzed independently. 
These motivated us to develop a more flexible information borrowing approach, which takes into account the potentially different similarity levels between indications. 
In addition to EXNEX and Liu's approach, there are other existing information borrowing approaches, 
such as BLAST (Chu and Yuan, 2018), 
which requires longitudinal biomarker measurements for promising performance, but the measurements are not always available in early-phase basket trials. Other approaches include 
Cunanan's two-path approach (Cunanan et al., 2017), 
Bayesian response-adaptive approach (Ventz et al., 2017), 
Chen's approach (Chen et al., 2016) 
and GSED (Magnusson and Turnbull, 2013), 
which either have similar ideas as EXNEX or Liu's approach, or show limited improvement in trial efficiency compared to independent analysis, or come with trials designs that are too complex for implementation, and therefore are not considered for further investigation.

In this paper, we first review several existing methods for the analysis of basket trials in early-phase oncology studies, including the independent analysis, BHM, EXNEX and Liu's approach. 
We then propose a novel approach using a Bayesian hierarchical model with a correlated prior (CBHM), which conducts more flexible borrowing based on the sample similarity between indications. 
CBHM introduces a correlation structure in the prior distribution of the log odds parameters that is determined by the sample ``distance'' between indications. 
This encourages information borrowing between the indications that appear to be homogeneous, and avoids excessive borrowing from the outliers, aiming to provide better type I error control while still improving the power of independent analysis. 
A two-stage trial design is proposed along with CBHM. 
We conducted simulation studies to investigate the relative performance between the various approaches discussed. 
Simulation results show that when the proportion of the sensitive indications is relatively large (e.g. $\geqslant 3$ out of 6), the information borrowing approaches (BHM, EXNEX, Liu's approach and CBHM) have substantially higher power than independent analysis. 
Our proposed CBHM approach shows advantage over the existing information borrowing approaches, with a robust overall performance, a power comparable to that of EXNEX or Liu's approach depending on the data scenario, and the potential to provide substantially better type I error control especially when the total number of indications and the proportion of the sensitive indications are large.
An additional feature of CBHM is that it comes with a detailed instruction for prior specification and trial design in a general setting, which is not provided by most existing methods and ensures the applicability of CBHM in a wider range of basket trials.

The remainder of the paper is organized as follows. In Section 2, we briefly review several representative methods for the analysis of early-phase basket trials, 
then introduce our CBHM approach. 
In Section 3, we present simulation results to discuss the performance of the various approaches. 
Conclusions and discussion of future work are summarized in Section 4. 

\section{Methods}\label{sec2}
Suppose that in an early-phase basket trial, a small number of patients with $I$ cancer types that share the same genomic or molecular aberration are enrolled. 
We denote the true response rate of the $i^{th}$ indication as $p_i$. For simple illustration, we set the standard of care (null) response rate to $q_0$, and the target response rate to $q_1$, for all indications. Our objective is to test whether or not each indication responds positively to the treatment:
\begin{align*}
H_0: p_i\leqslant q_0, \text{\hspace{2pt}} \text{  versus  } \text{\hspace{5pt}} H_a: p_i \geqslant q_1, \text{  for } 
i=1,2,\ldots,I. 
\end{align*}
In Section 2.1, we introduce the two reference methods, the Bayesian independent approach and BHM. In Section 2.2, we give a brief overview of the two recently proposed information borrowing approaches, EXNEX and Liu's approach, whose performance we would like to investigate.  
We then introduce our proposed CBHM approach in Section 2.3.

\subsection{Reference methods}\label{reference}
\subsubsection{Independent analysis}
Independent analysis is simple and widely used in clinical trials. Assume that the binary response of the $j^{th}$ patient in the $i^{th}$ indication group is $Y_{ij}$, where $Y_{ij}=1$ indicates positive response to the treatment and $Y_{ij}=0$ indicates no positive response.  
Here we consider a Bayesian independent approach with a beta-binomial model, where $Y_{ij} \stackrel{i.i.d.}{\sim} \text{Ber}(p_i)$, with $\text{Ber}(p)$ denoting a Bernoulli distribution with probability $p$, and each $p_i$ is assumed to have a flat prior,  $\text{Beta}(1, 1)$, where $\text{Beta}(\alpha, \beta)$ denotes the Beta distribution with shape paramters $\alpha$ and $\beta$. The hypothesis testing procedure will be introduced in the trial process in Section 2.3.3.

\subsubsection{Bayesian hierarchical model (BHM)}
Thall et al. (2003) and Berry et al. (2013) discussed the use of Bayesian hierarchical model (BHM) to improve the trial efficiency of basket trials by borrowing information across indications (Thall et al., 2013; Berry et al., 2013). 
This conventional BHM assumes that the log odds parameters of all indications are exchangeable and center around a common mean, specifically, $\{\theta_i = \log(p_i/(1-p_i))$, $i=1,2,\ldots,I\}$ follow a common prior, $N(\theta_0,\sigma^2)$. 
By assuming common mean $\theta_0$ and variance $\sigma^2$, BHM combines information from all indications. 

\subsection{Two recently published information borrowing approaches}\label{existing}

\subsubsection{EXNEX}
Neuenschwander et al. (2016) proposed an EXNEX method, which robustifies the conventional BHM that has a full EX assumption (Neuenschwander et al., 2016).  
EXNEX adopts a Bayesian hierarchical mixture model that allows each log odds parameter $\theta_i=\log(p_i/(1-p_i))$ to be either exchangeable with parameters of some indications or nonexchangeable with any other parameter. 
This raises two possibility for each indication $i$: 
(1) EX: $\theta_i|\mu_0,\sigma^2_0 \sim N(\mu_0,\sigma^2_0)$ with probability $\pi_i$ and EX parameters $\mu_0$ and $\sigma^2_0$;
(2) NEX: 
$\theta_i|\mu_i,\sigma^2_i \sim N(\mu_i,\sigma_i^2)$,  
with probability $1-\pi_i$ and indication-specific parameters $\mu_i$ and $\sigma^2_i$. 

\subsubsection{Liu's two-path approach}
Liu et al. (2017) proposed to conduct an interim homogeneity test to decide the necessity of information borrowing across indications (Liu et al, 2017). 
Figure 1 shows the flow chart of Liu's approach. Specifically, at decision point (a), 
a homogeneity test is conducted on $\{p_1,p_2,\ldots,p_I\}$ using Cochran's Q test. If the homogeneity hypothesis is rejected and it is concluded that at least one indication is different from others, then the trial proceeds to the ``heterogeneity path'' using Simon's two-stage design (Simon, 1989). 
But if there is not enough evidence to reject homogeneity, the trial then proceeds to the ``homogeneity path'': first, a futility analysis is conducted at decision point (b)  by calculating the Bayesian predictive power given $\hat{p}_{i,1} =r_{i,1}/n_{i,1}$ ($n_{i,j}$ and $r_{i,j}$ denote the number of patients and responders, respectively, for the $i^{th}$ indication in stage $j$). After assuming a $\text{Beta}(0.5, 0.5)$ prior on $p_i$, $r_{i,2}$ 
is simulated from its posterior distribution based on the beta-binomial model $(n_{i,1} + 0.5, r_{i,1} + 0.5, n_{i,1})$. 
The final response rate, $\hat{p_i} =(r_{i,1}+r_{i,2})/(n_{i,1}+n_{i,2})$, and the Bayesian predictive power of $(\hat{p_i} > p_0)$ are then calculated. If the predictive power is less than a threshold $C$, the enrollment will be stopped early, and the indication is concluded to be insensitive to the treatment. 
In the end, a BHMM with two mixture components is applied to the final data:
\begin{align*}
Y_{ij} \stackrel{i.i.d.}{\sim} & \text{Ber}(p_i),\nonumber\\
\theta_i = \log & \frac{p_i}{1-p_i}, \nonumber\\
\theta_i = \text{ } \delta_i M_{i1} & + (1-\delta_i) M_{i2},\nonumber\\
\delta_i\sim \text{Ber}(\pi_i), \text{  }
M_{i1} \sim N & (\mu_1,\sigma^2_1), \text{  } M_{i2} \sim N(\mu_2,\sigma^2_2),
\end{align*}
where $\delta_i$ is the indicator for the mixture component the $i^{th}$ indication belongs to, with $\delta_i=1$ indicating the first component and $\delta_i=0$ indicating the second, $\pi_i$ is the prior probability of $\delta_i=1$, $M_{ij}$ is the logit response rate of the $i^{th}$ indication if the indication belongs to the $j^{th}$ component, $\mu_j$ and $\sigma^2_j$ are the prior mean and variance parameters, respectively, of $M_{ij}$,  $i=1,2,\ldots,I$, $j=1,2$.

\begin{figure*}[ht!]
	\centering
    \includegraphics[width=300pt]{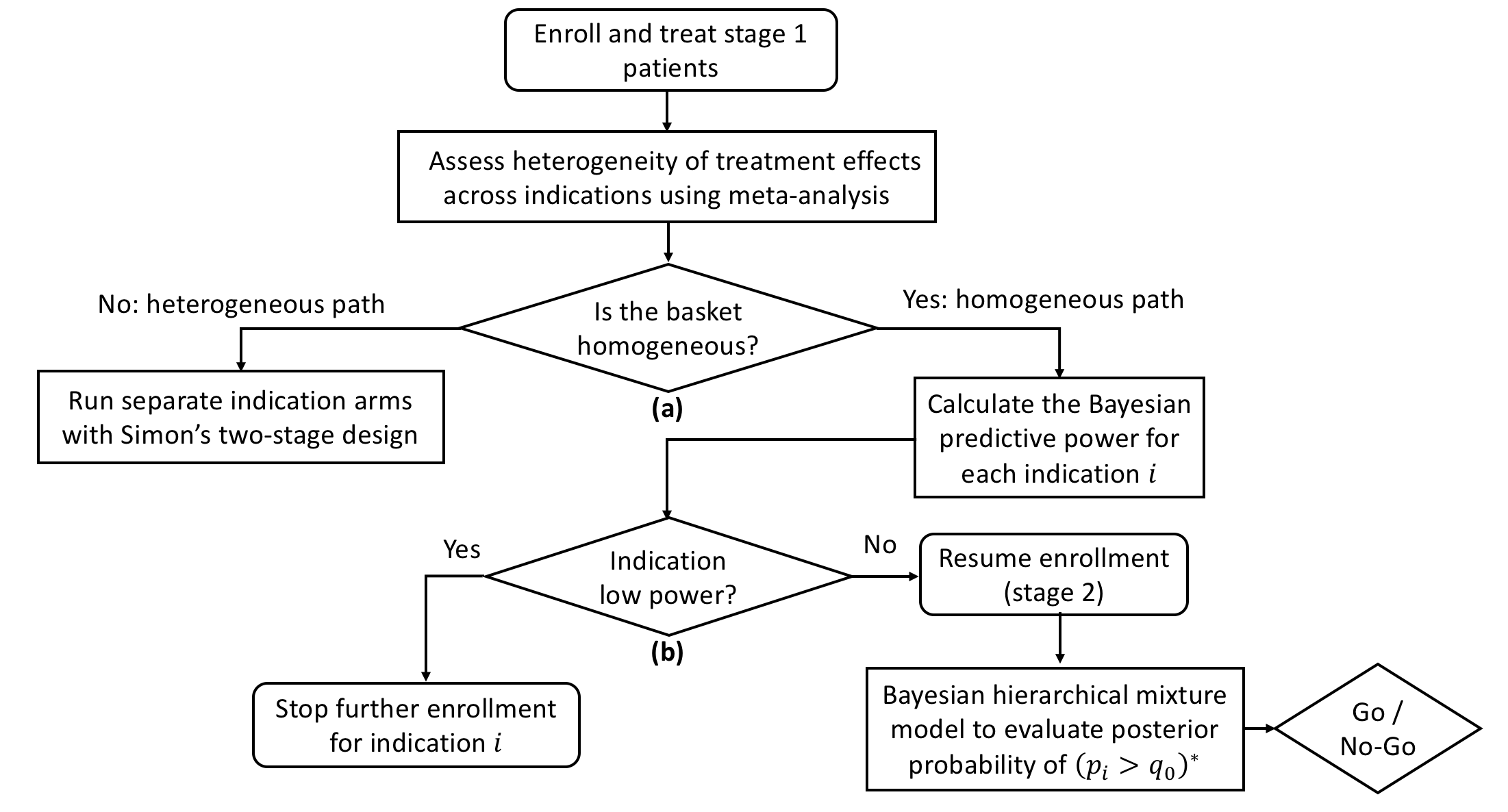}
    \caption{Flow chart of Liu's two-path approach, 
    $p_i$ denotes the response rate of the $i^{th}$ indication, and $q_0$ is the null response rate \protect (Liu et al., 2017). The indications that are concluded to be homogeneous but do not pass the predictive power evaluation at (b) are also included in the BHMM analysis.\label{liu_flowchart}}
\end{figure*}
\FloatBarrier
\noindent

\subsection{Bayesian hierarchical model with a correlated prior (CBHM)}\label{proposed}

\subsubsection{Model specification}\label{cbhmmodel}
The CBHM approach was motivated by 
the goal of borrowing more information between indications that are possibly homogeneous, and less information from the indication that is likely an outlier. 
The basic idea is to conduct flexible borrowing based on pairwise sample similarity of the indications, which can be described by the pairwise sample correlations.
We utilize the idea of modeling the spatial correlation between geostatistical locations: 
imagine that the indications are different locations on a map, and $d_{ij}$, the distance between locations $i$ and $j$, is defined as the distance between the posterior distributions of their response rates. 
To induce information borrowing based on sample similarity, we introduce correlated indication-specific effects, $\bm{\eta}=(\eta_1,\eta_2,\ldots,\eta_I)^T$, on the vector of log odds parameters, $\bm{\theta}=(\theta_1,\theta_2,\ldots,\theta_I)^T$, and assume that $\theta_i=\theta_0+\eta_i+\epsilon_i$, 
with $\theta_0$ being the common mean, and $\bm{\epsilon}=(\epsilon_1,\epsilon_2,\ldots,\epsilon_I)^T$ being a vector of independent random errors. 
The model setup for CBHM is then: 
\begin{align}
 & Y_{ij} \stackrel{i.i.d.}{\sim} \text{Ber}(p_i),\nonumber\\
\theta_i & = \log \frac{p_i}{1-p_i}, \nonumber\\
\theta_i & = \theta_0 + \eta_i + \epsilon_i, \nonumber\\
\bm{\eta}
\sim \mathcal{MVN}(\bm{0}, & \sigma^2 \bm{R}(\phi)),\text{  }
\epsilon_i \stackrel{i.i.d.}{\sim} N(0,\tau^2),
\label{cbhm}
\end{align}
where $\bm{R}(\phi)$ denotes the between-indication correlation matrix for $\bm{\theta}$, and the sample similarity between indications $i$ and $j$ is reflected by its $(i,j)$-th entry, $R_{ij}(\phi)=\rho(d_{ij}|\phi)$, 
$\sigma^2$ and $\tau^2$ denote the between- and within-indication variance, respectively. 
Model (1) is developed for the basket trials with binary outcome. For a trial with continuous outcome, the model can be modified by replacing the logit link function with a link function that transforms $Y_{ij}$ to be approximately normally distributed.

We now discuss the construction of the correlation matrix $\bm{R}(\phi)$. In early-phase oncology trials, the endpoints are often defined as the patients' binary responses, in which case it is infeasible to calculate the commonly used Pearson correlation between response rates. We instead define the correlation as a function of the sample ``distance'': the distance between the posterior distributions of the response rates.
There are multiple choices regarding the measure of distance between two probability distributions. One is the commonly used Kullback-Leibler (KL)  divergence, which measures the directed divergence from one distribution to another (Burnham and Anderson, 2001). 
We denote the data for the $i^{th}$ and $j^{th}$ indications as $D_i=\{n_i,r_i\}$ and $D_j=\{n_j,r_j\}$, respectively, where $n_i$ and $r_i$ are the sample size and the number of responders, respectively. The KL divergence from the posterior distribution of $p_j$, which we denote as $(p_j|D_j)$, to the posterior distribution of $p_i$, $(p_i|D_i)$, 
is defined as:
\begin{align*}
D_{KL}(f(p_i|D_i)||f(p_j|D_j))=\int f(p|D_i)\log\frac{f(p|D_i)}{f(p|D_j)}dp,
\end{align*}
where $f(p_i|D_i)$ denotes the probability density function (pdf) of $(p_i|D_i)$.
Note that the KL divergence from $(p_j|D_j)$ to $(p_i|D_i)$ is not equal to the KL divergence from $(p_i|D_i)$ to $(p_j|D_j)$. To define a symmetric distance measure, we take the average of the two and define the KL distance between the $i^{th}$ and $j^{th}$ indications as:
\begin{align*}
d^{KL}_{i,j}=\frac{1}{2}\Big(\int f(p|D_i)\log\frac{f(p|D_i)}{f(p|D_j)}dp + \int f(p|D_j)\log\frac{f(p|D_j)}{f(p|D_i)}dp\Big).
\end{align*}
There are symmetric distance measures as well, such as the Hellinger (H) distance (Hellinger, 1909): 
\begin{align*}
d^{H}_{i,j}=\sqrt[]{1-\int \sqrt[]{f(p|D_i)f(p|D_j)}dp},
\end{align*}
and the Bhattacharyya (B) distance (Kailath, 1967): 
\begin{align*}
d^{B}_{i,j}=-\log\Big\{\int \sqrt[]{f(p|D_i)f(p|D_j)}dp\Big\}.
\end{align*}
\begin{figure*}[htb]
\centering
\includegraphics[width=185pt]{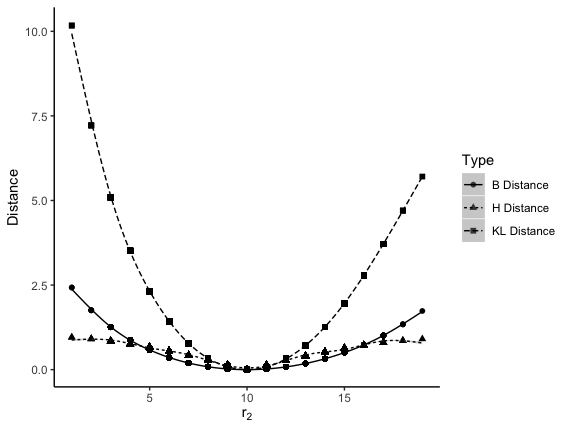}
    \caption{B, H and KL distance between two indications given sample sizes $n_1=n_2=24$, the number of responders $r_1=10$ and $r_2$ varying between $0$ and $n_2$. 
    }
\label{dist_plots}
\end{figure*}
In general, different distance measures vary by the smoothness and scale.  
Figure 2 shows the different behaviors of KL, H and B distance. 
We can observe that H distance is the most smooth, which changes slowly as the difference between sample response rates changes, and lies within $[0,1)$.
KL distance is the least smooth with a much larger differentiability and lies within $[0,+\infty)$. B distance is more smooth than KL distance and less smooth that H distance, and has a support of $[0,+\infty)$.

Given distance $d_{ij}$, the sample correlation between the $i^{th}$ and $j^{th}$ indications is defined as a function of $d_{ij}$ with a set of correlation parameters, $\bm{\zeta}$. 
There are many choices for the correlation function, among which 
one of the most widely used one is the mat\'ern correlation function: 
$\rho(d_{ij}|\bm{\zeta})=\frac{1}{2^{\nu-1} \Gamma({\nu})} \Big(\frac{2\nu^{1/2}d_{ij}}{\phi}\Big)^{\nu}\bm{J}_{\nu} \Big(\frac{2\nu^{1/2}d_{ij}}{\phi}\Big)
$, where $\bm{\zeta}=\{\phi,\nu\}$, $\phi$ is the range parameter (larger $\phi$ indicates larger-scale correlation), $\nu$ is the smoothness parameter (larger $\nu$ indicates smoother correlation), 
$\Gamma(\cdot)$ is the gamma function, and $\bm{J}_{\nu}(\cdot)$ is the modified Bessel function of the second kind with order $\nu$ (Stein, 2012). 
We consider two commonly used special cases of Mat\'ern correlation: exponential correlation ($\nu=0.5$) and squared exponential correlation ($\nu\rightarrow\infty$), where the former is defined as $e^{-\phi d_{ij}}$ and the latter is defined as $e^{-{\phi} d_{ij}^2}$. Figure 3 shows a comparison between the exponential and squared exponential correlation functions under different values of $\phi$. We can see that the squared exponential correlation is more smooth, 
and compared to the exponential correlation, it induces higher correlation (i.e. stronger borrowing) under small distance, and lower correlation (i.e. weaker borrowing) under large distance.
\begin{figure*}[htp!]
    \centering
    \includegraphics[width=130pt]{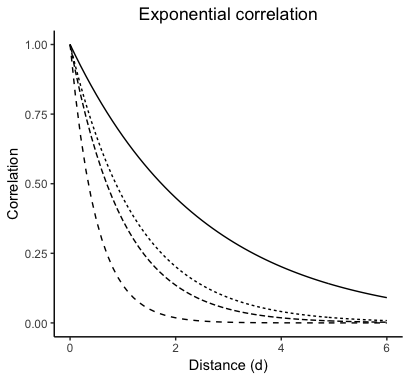}
    \centering\includegraphics[width=150pt]{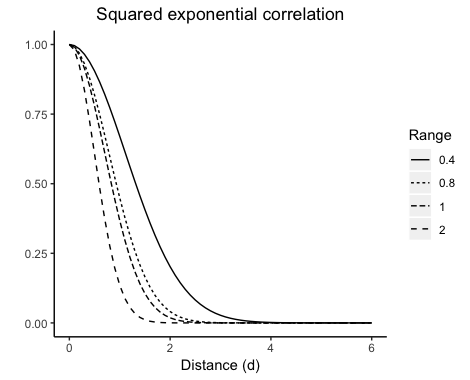}
    \caption{A comparison between the behaviors of exponential correlation function and squared exponential correlation function. The term ``range'' refers to the correlation range parameter $\phi$.}
\label{cor_plots}
\end{figure*}

\subsubsection{Bayesian inference}
We assign the following prior distribution to the model parameters:
\begin{align*}
\theta_0 \sim N(\mu_0,\sigma^2_0) &, \phi \sim \text{G}(a,1),\\
\sigma^2 \sim \text{IG}(c_{\sigma^2},d_{\sigma^2}),
\tau^2 \sim \text{IG} & (c_{\tau^2},d_{\tau^2}),
\sigma_0^2 \sim \text{IG}(c_{\sigma_0^2},d_{\sigma_0^2}),
\end{align*}
where $\text{G}(a,b)$ and $\text{IG}(a,b)$ denote the Gamma distribution and inverse Gamma distribution, respectively, with shape parameter $a$ and rate parameter $b$. 
We set the rate parameter in $\text{G}(a,1)$ to 1 considering that the rate parameter only affects the scale of the prior: if $x\sim \text{G}(a,1)$, then $\beta x \sim \text{G}(a,1/\beta)$. 
Note that in the early-phase basket trials, there is usually a limited sample size per indication, and thus for all Bayesian approaches, the prior specification will have certain impact on the posterior distribution. 
For CBHM, for example, smaller $a$ in the $\text{G}(a,1)$ prior for $\phi$ leads to stronger borrowing between indications. In practice, $a$ can be chosen by simulation based on $q_0$, $q_1$, distance measure, correlation function, etc. 
A formal instruction for the prior specification of CBHM is provided in Appendix D.

\subsubsection{Trial process}\label{trialdesign}
For the implementation of CBHM, we propose a two-stage design, which can be viewed as a special case of the BLAST design (Chu and Yuan, 2018): 
\begin{itemize}
\item[] 
\textit{\textbf{Step 1:}} in stage one, enroll $n_{i,1}$ patients for the $i^{th}$ indication group and collect data $D_1=\{n_{i,1},Y^1_{ij},i=1,\ldots,I,j=1,\ldots,n_{i,1}\}$, where $n_{i,1}$ denotes the stage-one sample size of the $i^{th}$ indication group.
\item[] 
\textit{\textbf{Step 2:}} at the end of stage one, apply model (1) to $D_1$ and conduct interim analysis on each indication $i$: if $Pr(p_i>(q_0+q_1)/2|D_1)<Q_f$, then stop enrollment early and conclude that the $i^{th}$ indication is not sensitive to the treatment; otherwise, continue to stage two until a total of $n_i$ patients 
are enrolled. 
\item[] 
\textit{\textbf{Step 3:}} at the end of the trial, assess the indication groups that continued to stage two: if $Pr(p_i>q_0|D)>Q$, where $D=\{n_{i},Y_{ij},i:n_{i,2}>0,j=1,\ldots,n_i\}$ denotes the final data, then the $i^{th}$ indication is concluded to be sensitive. 
\end{itemize}
Here we use $(q_0+q_1)/2$ as the threshold for response rate in the interim analysis considering that it has been used in other designs as well (Chu and Yuan, 2018). 
Simulation results also show that the choice for the threshold has little impact on the performance of CBHM. 
The probability cut-off for final decision, $Q$, can be calibrated to control the type I error rate (Thall et al., 1995; Chu and Yuan, 2018). 

\section{Simulation Studies}\label{simulation}
\subsection{Simulation settings}\label{simusetting}
We conducted simulation studies to investigate the relative performance between the various approaches in Section 2. 
We set the null response rate to $q_0=0.2$, target response rate to $q_1=0.4$, and assumed that there were $I=6$ indications that belonged to either a sensitive subgroup with response rate equal to $q_1$, or an insensitive subgroup with response rate equal to $q_0$. 
We considered various scenarios where there were $0,1,\ldots, I-1$  sensitive indications. 
For a fair comparison, we applied our two-stage trial design in Section 2.3.3 to all approaches except for Liu's approach, which should be implemented using its own design that is slightly different and hard to fit in our design (see Figure 1). 
For each indication, we set the maximum sample size to $n_i=24$, and conducted interim analysis when the number of enrolled patients reached $n^1_i=14$. This setting for $n_i$ and $n^1_i$ followed the minimax Simon's two-stage design with power $80.0\%$, type I error rate $10.0\%$, null response rate $0.2$ and target response rate $0.4$ (Simon, 1989). 
$Y_{ij}$'s were then simulated independently from $\text{Ber}(p_i)$.
We set the probability cut-off for interim analysis to $Q_f=0.05$, 
and calibrated $Q$, the probability cut-off for final decision, 
to control the type I error rate at $\alpha=10.0\%$ under the null scenario.

We conducted CBHM using B distance with exponential correlation. 
Prior specification for the various Bayesian models, as well as the specification of thresholds and tuning parameters for EXNEX and Liu's approach, are provided in Appendices A, B, C and D. 
We ran 5000 simulations under each scenario, and in each simulation, we approximated the Bayesian posterior distribution using 10000 MCMC iterations after a burn-in stage of 5000 iterations, which, based on our simulation results, guaranteed the convergence of MCMC for all implemented Bayesian models. The simulations were conducted in R with the Bayesian models being implemented using the R package ``rjags'' (Plummer, 2013). 
The average computation time per simulation is
approximately 0.6 seconds for independent analysis, 0.8 seconds for BHM, 0.2 seconds for Liu's heterogeneous path, 0.9 seconds for Liu's homogeneous path, 1.9 seconds for EXNEX, and 4.9 seconds for CBHM.
\begin{table*}[htb]
\footnotesize
\centering
\caption{
Simulation results assuming 0-2 sensitive indications out of 6. 
For CBHM, results were obtained using B distance with exponential correlation. 
\label{simu1}}	\begin{tabular}{cclrrrrrrcccc}
			\hline
			\multirow {2}{*}{Scenario} & \multirow {2}{*}{Method} &  & \multicolumn{6} {c}{Results for Cancer Type:} & \multirow {2}{*}{\makecell{Sample\\Size}} & \multirow {2}{*}{\% Perfect} & \multirow {2}{*}{\# TP} & \multirow {2}{*}{\# TN}\\
            \cline{4-9}
            & & & 1 & 2 & 3 & 4 & 5 & 6 & & & &\\
			\hline 
			1  &  & True RRs & 0.2 & 0.2 & 0.2 & 0.2 & 0.2 & 0.2 & & & 0 & 6\\
            (Null) & \multirow {2}{*}{Independent} & \% reject & 10.7&9.6&10.0&9.5&10.7&9.9 & \multirow {2}{*}{132.0} & \multirow {2}{*}{53.0} & \multirow {2}{*}{0.00} & \multirow {2}{*}{5.40}\\ 
			&  & \% stop & 20.2&19.8&20.1&20.2&19.6&19.8 & & & &\\ 
            & \multirow {2}{*}{BHM} & \% reject & 10.2&9.8&9.9&10.0&10.2&9.9 & \multirow {2}{*}{115.6} & \multirow {2}{*}{72.2} & \multirow {2}{*}{0.00} & \multirow {2}{*}{5.40}\\ 
			&  & \% stop & 47.3&47.2&47.6&47.3&47.5&47.3 & & & &\\   
			& \multirow {2}{*}{EXNEX} & \% reject & 10.6&9.4&10.1&9.5&10.6&9.8 & \multirow {2}{*}{125.0} & \multirow {2}{*}{64.5} & \multirow {2}{*}{0.00} & \multirow {2}{*}{5.40} \\ 
			&  & \% stop & 32.0&31.4&31.6&31.8&31.9&31.7 & & & &\\
            & \multirow {2}{*}{Liu's} & \% reject & 10.8&9.6&10.0&9.4&10.4&10.1 & \multirow {2}{*}{119.6} & \multirow {2}{*}{59.7} & \multirow {2}{*}{0.00} & \multirow {2}{*}{5.40}\\ 
			&  & \% stop & 40.6&40.3&40.7&41.0&40.8&40.2 & & & &\\
            & \multirow {2}{*}{CBHM} & \% reject & 10.0&10.0&9.8&10.4&9.7&9.7 & \multirow {2}{*}{119.9} & \multirow {2}{*}{63.4} & \multirow {2}{*}{0.00} & \multirow {2}{*}{5.40}\\
			&  & \% stop & 40.5&41.0&40.2&39.3&40.2&39.9 & & & &\\\\
            2 &  & True RRs & \textbf{0.4} & 0.2 & 0.2 & 0.2 & 0.2 & 0.2 & & & 1 & 5\\
            & \multirow {2}{*}{Independent} & \% reject & 82.4&10.5&9.9&10.0&10.1&9.8 & \multirow {2}{*}{134.2} & \multirow {2}{*}{47.8} & \multirow {2}{*}{0.82} & \multirow {2}{*}{4.50}\\ 
			&  & \% stop &  0.7&18.8&20.3&19.7&19.6&19.0 & & & &\\            & \multirow {2}{*}{BHM} & \% reject & 71.3&18.2&18.5&18.9&18.5&18.5 & \multirow {2}{*}{129.4} & \multirow {2}{*}{29.0} & \multirow {2}{*}{0.71} & \multirow {2}{*}{4.07}\\ 
			&  & \% stop & 7.4&27.8&28.0&27.3&28.0&27.5 & & & &\\
			& \multirow {2}{*}{EXNEX} & \% reject & 78.9&14.2&14.3&14.5&14.0&14.2 & \multirow {2}{*}{132.5} & \multirow {2}{*}{39.7} & \multirow {2}{*}{0.79} & \multirow {2}{*}{4.29} \\ 
			&  & \% stop & 2.4&22.1&22.5&22.9&22.7&22.1 & & & &\\
            & \multirow {2}{*}{Liu's} & \% reject & 80.7&13.9&13.1&13.5&13.2&13.5 & \multirow {2}{*}{125.2}  & \multirow {2}{*}{41.4} & \multirow {2}{*}{0.81} & \multirow {2}{*}{4.32}\\ 
			&  & \% stop & 3.4&36.3&37.4&36.6&37.5&36.9 & & & &\\
			& \multirow {2}{*}{CBHM} & \% reject &  79.1&14.2&14.3&14.1&15.0&14.9 & \multirow {2}{*}{126.5} & \multirow {2}{*}{36.2} & \multirow {2}{*}{0.79} & \multirow {2}{*}{4.28}\\ 
			&  & \% stop & 3.7&34.4&33.7&34.7&34.6&34.1 & & & &\\\\
            3 &  & True RRs & \textbf{0.4} & \textbf{0.4} & 0.2 & 0.2 & 0.2 & 0.2 & & & 2 & 4\\ 
            & \multirow {2}{*}{Independent} & \% reject & 81.3&81.2&10.2&9.6&9.8&9.8 & \multirow {2}{*}{135.8} & \multirow {2}{*}{43.8} & \multirow {2}{*}{1.62} & \multirow {2}{*}{3.62}\\ 
			&  & \% stop & 0.9&0.9&19.4&20.0&20.0&21.1 & & & &\\	            & \multirow {2}{*}{BHM} & \% reject & 83.4&83.5&27.1&25.8&26.7&25.4 & \multirow {2}{*}{137.0} & \multirow {2}{*}{26.8} & \multirow {2}{*}{1.67} & \multirow {2}{*}{2.95}\\ 
			&  & \% stop & 3.3&3.2&15.3&16.1&16.2&16.3 & & & &\\
			& \multirow {2}{*}{EXNEX} & \% reject & 84.3&84.0&18.5&18.1&18.8&17.7 & \multirow {2}{*}{136.6} & \multirow {2}{*}{33.1} & \multirow {2}{*}{1.68} & \multirow {2}{*}{3.27} \\ 
			&  & \% stop & 1.5&1.5&17.2&18.0&17.2&18.7 & & & &\\
            & \multirow {2}{*}{Liu's} & \% reject & 84.0&84.4&16.3&15.7&16.4&15.2 & \multirow {2}{*}{129.1}  & \multirow {2}{*}{37.1} & \multirow {2}{*}{1.68} & \multirow {2}{*}{3.37}\\ 
			&  & \% stop & 3.0&2.6&35.0&36.7&35.0&36.3 & & & &\\
		    & \multirow {2}{*}{CBHM} & \% reject & 85.1&85.7&16.9&16.7&16.5&15.9 & \multirow {2}{*}{131.6} & \multirow {2}{*}{33.4} & \multirow {2}{*}{1.71} & \multirow {2}{*}{3.33}\\ 
			&  & \% stop &  3.2&2.4&29.0&30.2&29.7&29.8 & & & &\\
			\hline
		\end{tabular}
\end{table*}

\subsection{Simulation results}\label{simuresults}
Tables 1-2 summarize the operating characteristics of the various approaches in different simulated scenarios.
The ``true RRs'' row lists the true response rate of each indication, with bold font indicating sensitivity to the treatment. The ``\% reject'' and ``\% stop'' row report the rejection percentage for $H_0$ and the early stopping percentage, respectively. The ``Sample Size'' column reports the average sample size considering potential early stopping. We also report the percentage of the simulations that gave correct conclusion for all indications (``\% perfect''), the average number of sensitive and insensitive indications each simulation correctly identified (``\# TP'' and ``\# TN'', respectively). 
We focus our discussion on the power, type I error control and average sample size.

In the null scenario in Table 1, the type I error rate is controlled at approximately $10\%$  for each approach, 
and compared to independent analysis, all information borrowing approaches 
tend to stop the trial early and thus have smaller average sample sizes.
In scenario 2 where only one indication is sensitive, 
EXNEX, Liu's approach and our proposed CBHM approach have slightly lower power ($78.9\%$, $80.7\%$ and $79.1\%$, respectively) than independent analysis ($82.4\%$), and slightly inflated average type I error rates ($14.2\%$, $13.4\%$ and $14.5\%$, respectively). 
BHM, on the other hand, has a considerably lower power (71.3\%) and higher average type I error rate ($18.5\%$). 
Liu's approach has the smallest average sample size, because its heterogeneous path uses Simon's two-stage design that has a more strict early stopping rule than the design for other approaches. 
CBHM has a similar average sample size as Liu's approach which is lower than that of the other approaches. 
In scenario 3 where there are 2 sensitive indications, 
all information borrowing approaches have higher power than independent analysis, among which CBHM has the highest average power (BHM: $83.5\%$, EXNEX: $84.2\%$, Liu's: $84.2\%$ and CBHM: $85.4\%$). 
EXNEX, Liu's approach and CBHM have less inflated average type I error rates ($18.5\%$, $15.9\%$ and $16.5\%$, respectively) than BHM ($26.3\%$).
BHM and EXNEX have the highest average sample sizes, 
while Liu's approach and CBHM have the lowest.

In Table 2 where 3-5 indications are sensitive, Liu's approach has the smallest average sample size because of its more strict early stopping rule, \begin{table*}[htb]
\footnotesize
\centering
\caption{
Simulation results assuming 3-5 sensitive indications out of 6. 
For CBHM, results were obtained using B distance with exponential correlation. \label{simu2}}
	\begin{tabular}{cclrrrrrrcccc}
			\hline
			\multirow {2}{*}{Scenario} & \multirow {2}{*}{Method} &  & \multicolumn{6} {c}{Results for Cancer Type:} & \multirow {2}{*}{\makecell{Sample\\Size}} & \multirow {2}{*}{\% Perfect} & \multirow {2}{*}{\# TP} & \multirow {2}{*}{\# TN}\\
            \cline{4-9}
            & & & 1 & 2 & 3 & 4 & 5 & 6 & & & &\\
			\hline 
            4 &  & True RRs & \textbf{0.4} & \textbf{0.4} & \textbf{0.4} & 0.2 & 0.2 & 0.2 & & & 3 & 3\\
            & \multirow {2}{*}{Independent} & \% reject & 81.6&81.0&81.3&10.0&10.0&9.8 & \multirow {2}{*}{137.8} & \multirow {2}{*}{39.8} & \multirow {2}{*}{2.44} & \multirow {2}{*}{2.70}\\ 
			&  & \% stop &0.8&0.8&0.8&19.5&19.9&19.8 & & & &\\	
            & \multirow {2}{*}{BHM} & \% reject & 89.9&89.7&90.5&36.3&35.7&35.0 & \multirow {2}{*}{141.3} & \multirow {2}{*}{25.6} & \multirow {2}{*}{2.70} & \multirow {2}{*}{1.93}\\ 
			&  & \% stop & 1.3&1.0&1.2&8.2&7.6&7.7 & & & &\\			& \multirow {2}{*}{EXNEX} & \% reject & 88.0&87.7&88.3&23.8&22.8&22.3 & \multirow {2}{*}{139.6} & \multirow {2}{*}{33.1} & \multirow {2}{*}{2.64} & \multirow {2}{*}{2.31}\\ 
			&  & \% stop & 0.7&0.9&0.9&13.7&13.7&13.6 & & & &\\
            & \multirow {2}{*}{Liu's} & \% reject & 86.7&85.8&86.3&20.4&20.2&19.6 & \multirow {2}{*}{132.8} & \multirow {2}{*}{34.9} & \multirow {2}{*}{2.59} & \multirow {2}{*}{2.40}\\  
			&  & \% stop & 2.8&3.1&2.6&35.0&34.7&34.2 &  &  &  &\\
		    & \multirow {2}{*}{CBHM} & \% reject & 88.9&88.5&88.7&20.2&19.2&18.8 & \multirow {2}{*}{136.3} & \multirow {2}{*}{37.4} & \multirow {2}{*}{2.66} & \multirow {2}{*}{2.42}\\
			&  & \% stop & 2.2&2.3&2.2&24.3&23.1&23.3 & & & &\\\\ 
            5 &  & True RRs & \textbf{0.4} & \textbf{0.4} & \textbf{0.4} & \textbf{0.4} & 0.2 & 0.2 & & & 4 & 2\\
            & \multirow {2}{*}{Independent} & \% reject & 81.1&81.3&81.8&82.5&9.8&10.1 & \multirow {2}{*}{139.6} & \multirow {2}{*}{35.9} & \multirow {2}{*}{3.27} & \multirow {2}{*}{1.80}\\ 
			&  & \% stop & 1.0&0.9&1.0&0.9&20.0&20.0 & & & &\\
            & \multirow {2}{*}{BHM} & \% reject & 93.4&93.7&94.0&94.1&46.0&46.7 & \multirow {2}{*}{143.1} & \multirow {2}{*}{25.7} & \multirow {2}{*}{3.75} & \multirow {2}{*}{1.07}\\ 
			&  & \% stop & 0.4&0.4&0.5&0.4&3.7&4.0 & & & &\\
	       	& \multirow {2}{*}{EXNEX} & \% reject & 91.1   &  91.3   &  91.6  &  92.0   &  28.0   &  28.9 & \multirow {2}{*}{141.7} & \multirow {2}{*}{35.3} & \multirow {2}{*}{3.66} & \multirow {2}{*}{1.43}\\ 
			&  & \% stop & 0.6&0.6&0.7&0.7&10.4&10.2 & & & &\\
            & \multirow {2}{*}{Liu's} & \% reject & 87.8&87.4&87.6&89.0&24.5&24.6 & \multirow {2}{*}{135.7} & \multirow {2}{*}{32.1} & \multirow {2}{*}{3.52} & \multirow {2}{*}{1.51}\\
			&  & \% stop & 3.1&2.9&3.1&2.4&35.8&35.2 &  &  &  &\\
			& \multirow {2}{*}{CBHM} & \% reject & 90.1&90.3&90.7&91.1&20.9&21.8 & \multirow {2}{*}{139.8} & \multirow {2}{*}{41.6} & \multirow {2}{*}{3.59} & \multirow {2}{*}{1.58}\\ 
			&  & \% stop &1.6&1.5&1.7&1.6&17.8&17.6 & & & &\\\\ 
           6 &  & True RRs & \textbf{0.4} & \textbf{0.4} & \textbf{0.4} & \textbf{0.4} & \textbf{0.4} & 0.2 & & & 5 & 1\\
            & \multirow {2}{*}{Independent} & \% reject &  81.7&83.0&82.0&81.6&80.7&10.3 & \multirow {2}{*}{141.6} & \multirow {2}{*}{32.3} & \multirow {2}{*}{4.09} & \multirow {2}{*}{0.90}\\ 
			&  & \% stop & 0.8&0.8&0.9&0.8&0.9&19.7& & & &\\
            & \multirow {2}{*}{BHM} & \% reject & 96.7&96.9&97.0&96.7&96.6&63.1 & \multirow {2}{*}{143.8} & \multirow {2}{*}{25.7} & \multirow {2}{*}{4.84} & \multirow {2}{*}{0.37}\\ 
			&  & \% stop & 0.1&0.2&0.1&0.1&0.1&1.0 & & & &\\
			& \multirow {2}{*}{EXNEX} & \% reject & 93.6&93.6&93.9&93.6&93.3&33.6 & \multirow {2}{*}{143.2} & \multirow {2}{*}{45.5} & \multirow {2}{*}{4.68} & \multirow {2}{*}{0.66}\\ 
			&  & \% stop & 0.3&0.4&0.4&0.3&0.3&6.8 & & & &\\
            & \multirow {2}{*}{Liu's} & \% reject & 89.2&89.9&90.1&89.6&89.3&32.0 & \multirow {2}{*}{139.0} & \multirow {2}{*}{35.8} & \multirow {2}{*}{4.48} & \multirow {2}{*}{0.68}\\
			&  & \% stop & 2.7&3.0&2.9&3.0&2.9&36.0 &  &  &  &\\
			& \multirow {2}{*}{CBHM} & \% reject & 91.1&91.6&91.2&90.8&90.6&25.6 & \multirow {2}{*}{142.4} & \multirow {2}{*}{43.6} & \multirow {2}{*}{4.55} & \multirow {2}{*}{0.74}\\ 
			&  & \% stop & 0.8&1.1&1.1&0.9&1.2&10.7 & & & &\\
            \hline
		\end{tabular}
\end{table*}
but overall all approaches have similar average sample sizes. Compared to scenarios 2-3, all information borrowing approaches now show more advantage in power compared to independent analysis. 
When 3 indications are sensitive, BHM has a substantially higher average power than independent analysis ($90.0\%$ versus $81.3\%$), but also a highly inflated average type I error rate ($35.7\%$). EXNEX and Liu's approach have lower average power ($88.0\%$ and $86.3\%$, respectively) and lower average type I error rates ($23.0\%$ and $20.1\%$, respectively). 
CBHM has an similar average power as EXNEX (88.7\%), and the best controlled type I error rate among the information borrowing approaches (average: 19.4\%). 
\begin{table*}[htb!]
\footnotesize
\centering
\caption{
Simulation results of CBHM using different distance measures. ``CBHM-B'', ``CBHM-KL'' and ``CBHM-H'' denote the CBHM using B, 
KL and H distance, respectively. 
\label{distoctable1_supp}}
	\begin{tabular}{cclrrrrrrcccc}
			\hline
			\multirow {2}{*}{Scenario} & \multirow {2}{*}{Method} &  & \multicolumn{6} {c}{Results for Cancer Type:} & \multirow {2}{*}{\makecell{Sample\\Size}} & \multirow {2}{*}{\% Perfect} & \multirow {2}{*}{\# TP} & \multirow {2}{*}{\# TN}\\
            \cline{4-9}
            & & & 1 & 2 & 3 & 4 & 5 & 6 & & & &\\
			\hline 
			1 &  & True RRs & 0.2 & 0.2 & 0.2 & 0.2 & 0.2 & 0.2 & & & 0 & 6\\
            & \multirow {2}{*}{CBHM-B} & \% reject & 10.0&10.0&9.8&10.4&9.7&9.7 & \multirow {2}{*}{119.9} & \multirow {2}{*}{63.4} & \multirow {2}{*}{0.00} & \multirow {2}{*}{5.40}\\
			&  & \% stop & 40.5&41.0&40.2&39.3&40.2&39.9 & & & &\\
            & \multirow {2}{*}{CBHM-KL} & \% reject & 10.5&9.4&10.1&9.3&10.6&10.0 & \multirow {2}{*}{123.6} & \multirow {2}{*}{54.6} & \multirow {2}{*}{0.00} & \multirow {2}{*}{5.31}\\ 
			&  & \% stop & 37.4&37.3&37.9&37.5&37.0&36.9 & & & &\\            & \multirow {2}{*}{CBHM-H} & \% reject & 9.5&10.3&10.1&10.9&9.3&9.6 & \multirow {2}{*}{121.9} & \multirow {2}{*}{62.6} & \multirow {2}{*}{0.00} & \multirow {2}{*}{5.40}\\ 
			&  & \% stop & 37.0&37.3&36.3&34.4&38.8&37.2 & & & &\\\\   
            2 &  & True RRs & \textbf{0.4} & 0.2 & 0.2 & 0.2 & 0.2 & 0.2 & & & 1 & 5\\
            & \multirow {2}{*}{CBHM-B} & \% reject &  79.1&14.2&14.3&14.1&15.0&14.9 & \multirow {2}{*}{126.5} & \multirow {2}{*}{36.2} & \multirow {2}{*}{0.79} & \multirow {2}{*}{4.28}\\ 
			&  & \% stop & 3.7&34.4&33.7&34.7&34.6&34.1 & & & &\\
            & \multirow {2}{*}{CBHM-KL} & \% reject & 82.5&12.5&12.4&12.3&12.7&12.8 & \multirow {2}{*}{127.3} & \multirow {2}{*}{43.4} & \multirow {2}{*}{0.83} & \multirow {2}{*}{4.37}\\
			&  & \% stop &  3.2&32.8&33.1&32.6&32.8&31.9 & & & &\\            & \multirow {2}{*}{CBHM-H} & \% reject &  79.8&13.8&14.1&13.8&14.5&14.6 & \multirow {2}{*}{129.8} & \multirow {2}{*}{39.2} & \multirow {2}{*}{0.80} & \multirow {2}{*}{4.29}\\ 
			&  & \% stop & 2.9&27.8&27.7&28.2&28.0&27.3 & & & &\\\\ 
            3 &  & True RRs & \textbf{0.4} & \textbf{0.4} & 0.2 & 0.2 & 0.2 & 0.2 & & & 2 & 4\\ 
		    & \multirow {2}{*}{CBHM-B} & \% reject & 85.1&85.7&16.9&16.7&16.5&15.9 & \multirow {2}{*}{131.6} & \multirow {2}{*}{33.4} & \multirow {2}{*}{1.71} & \multirow {2}{*}{3.33}\\ 
			&  & \% stop &  3.2&2.4&29.0&30.2&29.7&29.8 & & & &\\
            & \multirow {2}{*}{CBHM-KL} & \% reject & 83.3&83.2&14.3&13.6&14.4&13.7 & \multirow {2}{*}{132.2} & \multirow {2}{*}{37.8} & \multirow {2}{*}{1.67} & \multirow {2}{*}{3.44}\\ 
			&  & \% stop & 2.6&2.5&27.7&28.3&28.2&29.2 & & & &\\ 
            & \multirow {2}{*}{CBHM-H} & \% reject & 85.5&85.4&17.0&16.1&17.4&16.9 & \multirow {2}{*}{135.2} & \multirow {2}{*}{35.8} & \multirow {2}{*}{1.71} & \multirow {2}{*}{3.33}\\
			&  & \% stop & 2.0&2.2&20.5&21.6&20.9&20.8 & & & &\\\\
            4 &  & True RRs & \textbf{0.4} & \textbf{0.4} & \textbf{0.4} & 0.2 & 0.2 & 0.2 & & & 3 & 3\\ 
            & \multirow {2}{*}{CBHM-B} & \% reject & 88.6&88.8&88.7&19.2&20.2&18.8 & \multirow {2}{*}{136.3} & \multirow {2}{*}{37.4} & \multirow {2}{*}{2.66} & \multirow {2}{*}{2.42}\\
			&  & \% stop & 2.2&2.3&2.2&24.3&23.1&23.3 & & & &\\
            & \multirow {2}{*}{CBHM-KL} & \% reject & 85.3&84.7&85.4&16.0&16.2&15.2 & \multirow {2}{*}{136.6} & \multirow {2}{*}{37.6} & \multirow {2}{*}{2.55} & \multirow {2}{*}{2.53}\\ 
			&  & \% stop &2.0&2.0&1.9&22.7&22.7&22.8& & & &\\       
            & \multirow {2}{*}{CBHM-H} & \% reject & 88.4&87.9&88.2&19.8&20.3&19.0 & \multirow {2}{*}{139.3} & \multirow {2}{*}{36.3} & \multirow {2}{*}{2.65} & \multirow {2}{*}{2.41}\\ 
			&  & \% stop &1.2&1.3&1.3&14.8&14.2&13.8& & & &  \\\\
            5 &  & True RRs & \textbf{0.4} & \textbf{0.4} & \textbf{0.4} & \textbf{0.4} & 0.2 & 0.2 & & & 4 & 2\\ 
			& \multirow {2}{*}{CBHM-B} & \% reject & 90.1&90.3&91.1&90.7&20.9&21.8 & \multirow {2}{*}{139.8} & \multirow {2}{*}{41.6} & \multirow {2}{*}{3.59} & \multirow {2}{*}{1.58}\\ 
			&  & \% stop &1.6&1.5&1.7&1.6&17.8&17.6 & & & &\\
            & \multirow {2}{*}{CBHM-KL} & \% reject & 86.4&86.2&86.6&87.7&17.0&17.1 & \multirow {2}{*}{139.8} & \multirow {2}{*}{38.3} & \multirow {2}{*}{3.47} & \multirow {2}{*}{1.66}\\ 
			&  & \% stop & 1.6&1.5&1.5&1.1&17.6&18.3 & & & &\\                & \multirow {2}{*}{CBHM-H} & \% reject & 89.3&89.9&90.6&89.9&24.6&24.1 & \multirow {2}{*}{141.8} & \multirow {2}{*}{38.0} & \multirow {2}{*}{3.60} & \multirow {2}{*}{1.51}\\  
			&  & \% stop & 0.7&0.6&0.6&0.7&10.0&9.7 & & & &\\\\
			6 &  & True RRs & \textbf{0.4} & \textbf{0.4} & \textbf{0.4} & \textbf{0.4} & \textbf{0.4} & 0.2 & & & 5 & 1\\
            & \multirow {2}{*}{CBHM-B} & \% reject & 91.1&91.6&91.2&90.8&90.6&25.6 & \multirow {2}{*}{142.4} & \multirow {2}{*}{43.6} & \multirow {2}{*}{4.55} & \multirow {2}{*}{0.74}\\ 
			&  & \% stop & 0.8&1.1&1.1&0.9&1.2&10.7 & & & &\\
            & \multirow {2}{*}{CBHM-KL} & \% reject & 88.6 & 89.0 &88.8 &88.1&87.9 &18.7 & \multirow {2}{*}{142.6} & \multirow {2}{*}{48.8} & \multirow {2}{*}{4.52} & \multirow {2}{*}{0.81}\\
			&  & \% stop & 0.1&0.6&0.9&0.8&0.7&10.4 & & & &\\                 & \multirow {2}{*}{CBHM-H} & \% reject & 92.0&92.2&92.4&91.6&91.7&30.5 & \multirow {2}{*}{143.2} & \multirow {2}{*}{41.2} & \multirow {2}{*}{4.60} & \multirow {2}{*}{0.69}\\ 
			&  & \% stop & 0.3&0.3&0.4&0.4&0.4&6.0 & & & &  \\    
			\hline
		\end{tabular}
\end{table*}
Results are similar when 4-5 indications are sensitive, and we observe a more clear advantage of CBHM over other information borrowing approaches especially in terms of type I error control: 
although BHM and EXNEX provide higher power than CBHM, the highly inflated type I error rates make their conclusions possibly unreliable. CBHM, on the contrary, provides a higher power than Liu's approach and substantially better control of type I error than all other information borrowing approaches.

We also summarized the absolute bias and root mean square error (RMSE) of the estimated response rates on Web Appendix.
Briefly, all approaches have similar absolute bias and RMSE. 
Liu's approach has the largest average absolute bias and RMSE in all studied scenarios except for the null scenario, where independent analysis has slightly larger average absolute bias and RMSE. Among information borrowing approaches, BHM, EXNEX and Liu's approach tend to have large absolute bias and RMSE for the outliers (i.e. the indications whose response rates differ from those of the majorities). CBHM, however, has more robust estimation with smaller absolute bias and RMSE for the outliers, as well as smaller abolute bias and RMSE than Liu's approach in all studied scenarios.

\subsection{Sensitivity Analysis}\label{sensitivity_analysis}
The simulation results of CBHM in Tables 1-2 were obtained using B distance. 
We also conducted simulations using KL and H distance to evaluate the sensitivity of CBHM with respect to the choice for distance measure. The operating characteristics of CBHM using different distance measures are summarized in Table 3, from which we can observe that the B distance and H distance lead to similar performance of CBHM, as well as similar relative performance between CBHM and other considered approaches. 
The KL distance, on the other hand, 
\begin{figure*}[htp]
\centerline{
\includegraphics[width=300pt]{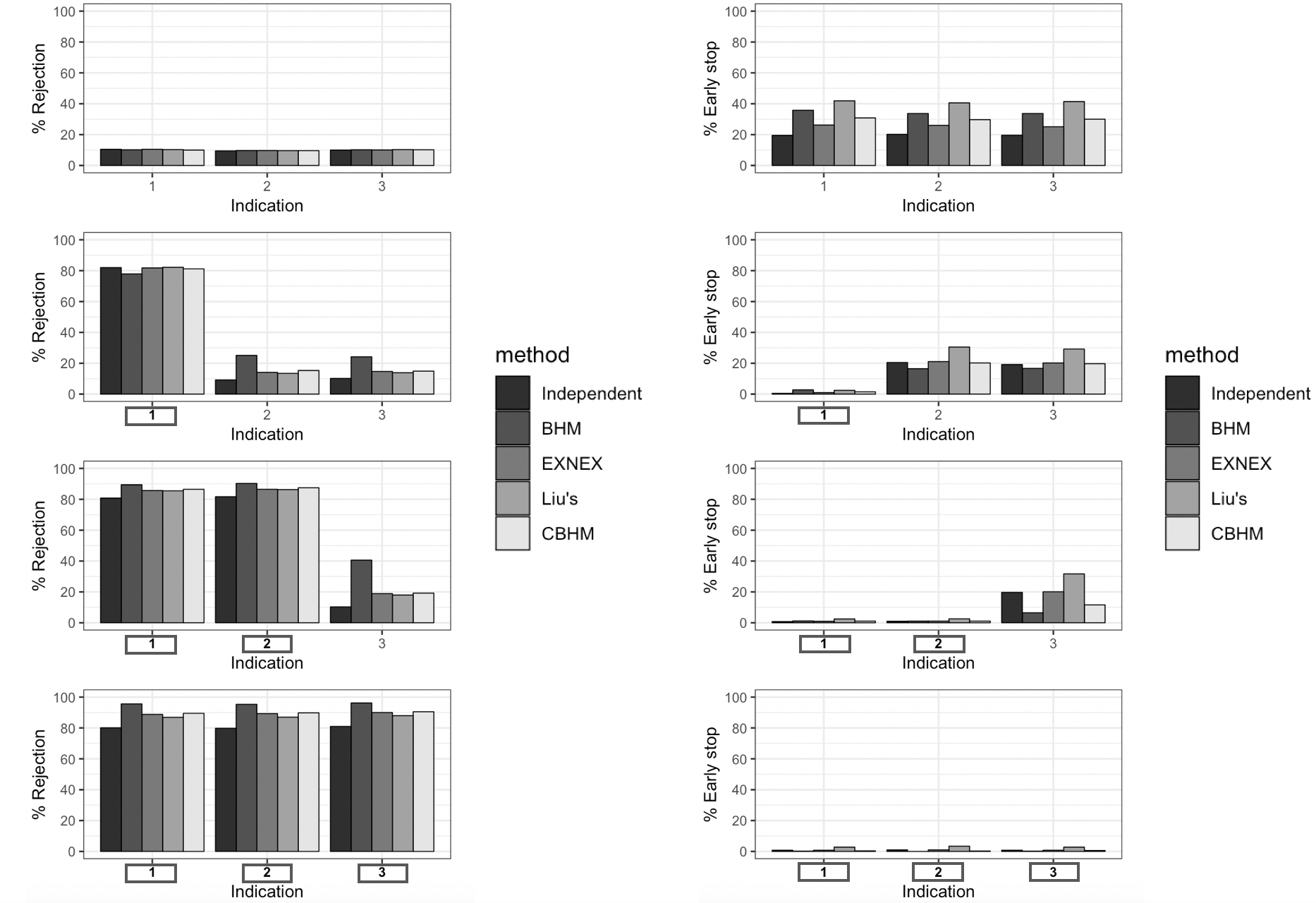}
}
\caption{The rejection percentage (i.e. the power for the sensitive indications and type I error rate for the insensitive indications, sub-figure 1) and early stopping percentage at interim analysis (sub-figure 2) assuming there are $I=3$ indications. 
CBHM was implemented using B distance with exponential correlation. The indications whose indexes on the x-axis are circled and marked in bold are sensitive to the targeted treatment, while the others are insensitive.\label{k3}}
\end{figure*}
tends to have lower power, more strict type I error control, and requires more complex procedure for prior specification. 
Note that different distance measures need to be implemented along with different correlation functions and priors. 
A detailed instruction and additional simulation results are provided on Web Appendix.
In practice, we recommend B distance with exponential correlation function given the promising performance and simple procedure required for prior specification.

\begin{figure*}[htb!]
\centering
\includegraphics[width=420pt]{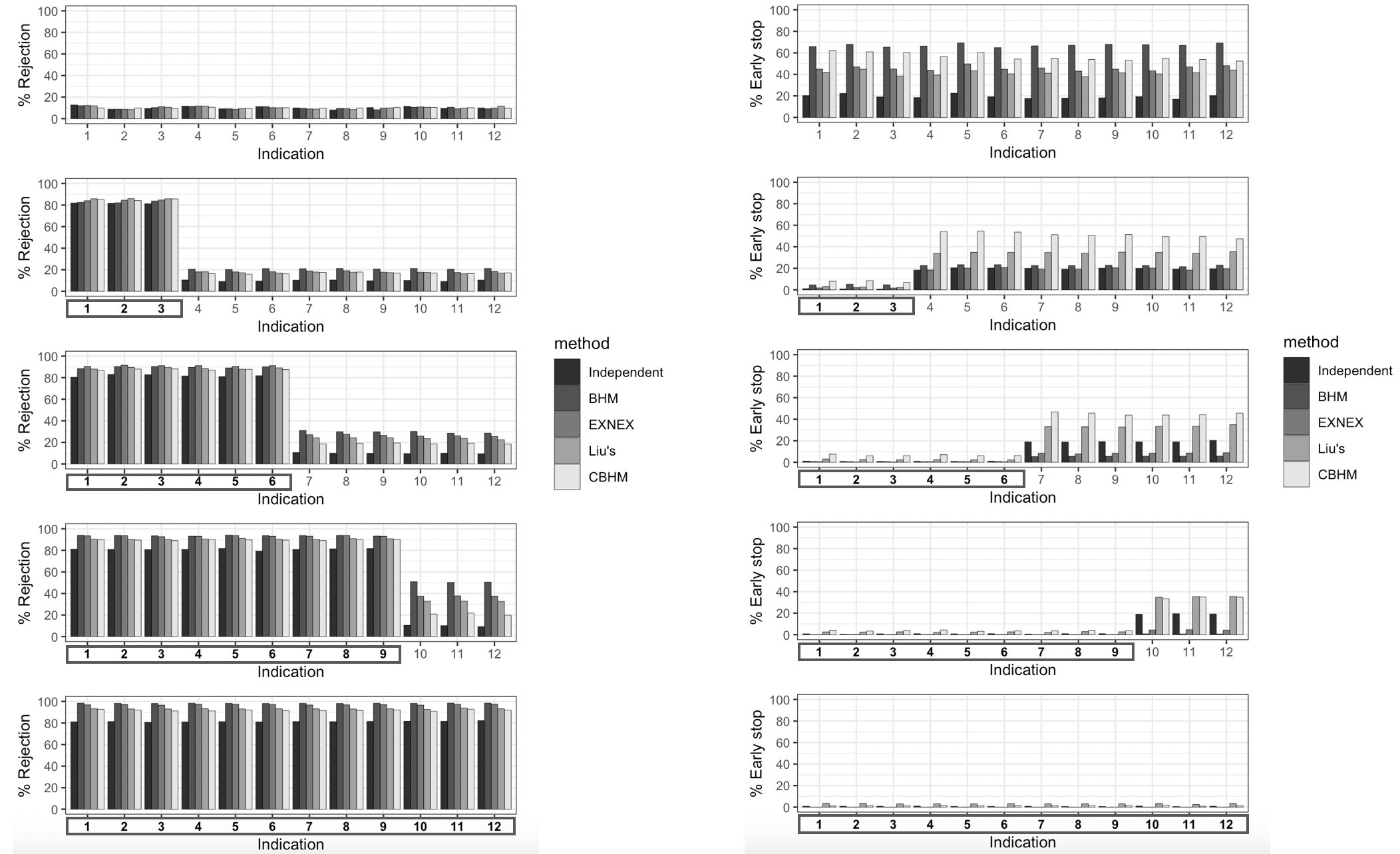}
\caption{Rejection percentage (i.e. power for the sensitive indications and type I error rate for the insensitive indications, sub-figure 1) and early stopping percentage at interim analysis (sub-figure 2) assuming there are $I=12$ indications. 
The indications whose indexes on the x-axis are circled and marked in bold are sensitive to the targeted treatment, while the others are insensitive.}
\label{k12}
\end{figure*}

So far, we have discussed simulation studies assuming a total of $I=6$ indications. We also investigated the performance of the various approaches when $I=3$ and 12. The rejection and early stopping percentages are reported in Figures 4-5. 
When $I=3$, EXNEX, Liu's approach and CBHM have similar performance, with lower power and more strict type I error control than BHM. 
When $I=12$, BHM and EXNEX show potential advantage in power but only when a large proportion of the indications (e.g. $\geqslant 9$ out of 12) are truly sensitive. They also have poor type I error control and low early stopping percentage. 
The advantage of CBHM is highlighted, with a promising power similar to Liu's approach, substantially better control of type I error rate compared to BHM, EXNEX and Liu's approach, and a high early stopping percentage. 
Overall, as the total number of indications increases, the difference  between the various approaches and the advantage of CBHM become more clear.

\begin{table*}[htb]
\centering
\caption{The various priors considered for $\sigma^2, \tau^2$ and $\phi$ in sensitivity analysis. 
\label{prior_specification}}
	\begin{tabular}{ccc}
			\hline
            Prior Setting & $\sigma^2, \tau^2$ & $\phi$\\
			\hline 
			1 & $\text{IG}(0.1,0.1)$ & $\text{G}(1,1)$\\
            2 & $\text{IG}(0.01,0.01)$ & $\text{G}(1,1)$\\
			3 & $\text{IG}(0.001,0.001)$ & $\text{G}(1,1)$\\
            4 & $\text{IG}(0.01,0.01)$ & $\text{G}(0.7,1)$\\
			\hline
		\end{tabular}
\end{table*}
We also conducted sensitivity analysis for the prior specification of CBHM. 
The prior for the between- and within-indication variance parameters ($\sigma^2$ and $\tau^2$, respectively) can have impact on the posterior distribution due to limited data. 
We therefore considered two other priors: $\text{IG}(0.1,0.1)$ and $\text{IG}(0.001,0.001)$, in addition to the $\text{IG}(0.01,0.01)$ prior in Section 3.2, for $\sigma^2$ and $\tau^2$. 
Following our instruction for prior specification in Appendix D, a $\text{G}(a,1)$ prior with $a\in [0.70,1.21]$ is reasonable for the correlation range parameter, $\phi$. 
We therefore considered another $\text{G}(0.7,1)$ prior as a comparison with the $\text{G}(1,1)$ prior in Section 3.2.  
Table 4 summarizes the four prior settings we considered for $\{\sigma^2,\tau^2,\phi\}$, while the prior for the rest of the model parameters remains the same as that in Section 3.2. 
Table 5 summarizes the simulation results. 
Overall, different priors for $\sigma^2$ and $\tau^2$ lead to similar performance of CBHM, but a $\text{IG}(c,c)$ prior with smaller $c$ tends to induce stronger borrowing that leads to higher power and type I error rate. We recommend the $\text{IG}(0.01,0.01)$ prior for application given its promising performance, but different priors can be considered depending on the primary focus, e.g. higher power or more strict type I error control.
Regarding the two priors for $\phi$ that were both selected based on our instruction for prior specification, $\text{G}(0.7,1)$ leads to slightly higher power and type I error rate than $\text{G}(1,1)$, but overall the results are similar. This suggests that the performance of CBHM is consistent under different choices for $a$ in the $\text{G}(a,1)$ prior for $\phi$, if $a$ is selected according to our instruction.

We also conducted a simulation study assuming more moderate treatment effect with $q_1=0.3$ instead of 0.4. Simulation results are summarized on Web Appendix, from which we observe a similar pattern of the relative performance between approaches as when $q_1=0.4$. When 0-2 of the 6 indications are sensitive, all approaches have similar and more promising performance than BHM in terms of power and type I error control. When 3-5 of the 6 indications are sensitive, CBHM has a power lower than BHM and EXNEX but higher than Liu's approach, and a similar type I error rate as Liu's approach that is much lower than that of BHM and EXNEX.

As a summary, Table 6 lists a comparison between the various approaches in terms of power, type I error control and design complexity. Regarding design complexity, CBHM avoids the need to conduct subgroup clustering required for EXNEX and the heterogeneity test required for Liu's approach, therefore has fewer tuning parameters and a simpler design. 
BHM is the least recommended due to the potentially huge inflation in type I error rate. 
Overall, the independent analysis has the advantage of strong type I error control and easy implementation, and is recommended when 
there is a priori information that only a small proportion (e.g. 1 out of 6) of the indications might be sensitive to the treatment. 
Our CBHM approach has the advantage of high detection power and potentially much better type I error control than BHM, EXNEX and Liu's approach. It is especially recommended when there is a relatively large number of indications in the basket (e.g. $I>3$), and when prior knowledge suggests that a relatively large proportion of the indications are sensitive to the treatment.

\begin{table*}[htb!]
\footnotesize{}
\centering
\caption{ 
Operating characteristics of CBHM using different priors for $\phi$, $\sigma^2$ and $\tau^2$. 
\label{sensitivity-prior}}
	\begin{tabular}{cclrrrrrrcccc}
			\hline
			\multirow {2}{*}{Scenario} & \multirow {2}{*}{Design} &  & \multicolumn{6} {c}{Results for Cancer Type:} & \multirow {2}{*}{\makecell{Sample\\Size}} & \multirow {2}{*}{\% Perfect} & \multirow {2}{*}{\# TP} & \multirow {2}{*}{\# TN}\\
            \cline{4-9}
            & & & 1 & 2 & 3 & 4 & 5 & 6 & & & &\\
			\hline 
			1 &  & True RRs & 0.2 & 0.2 & 0.2 & 0.2 & 0.2 & 0.2 & & & 0 & 6\\
            & \multirow {2}{*}{Prior 1} & \% reject & 10.3 & 9.9 & 10.6 & 9.6 & 9.9 & 9.9  & \multirow {2}{*}{121.1} & \multirow {2}{*}{57.7} & \multirow {2}{*}{0.00} & \multirow {2}{*}{5.38}\\ 
			&  & \% stop & 38.2&39.0  &   38.3  &  37.0  &   38.5  &   38.2 & & & &\\ 
            & \multirow {2}{*}{Prior 2} & \% reject &  10.0&10.0&9.8&10.4&9.7&9.7  & \multirow {2}{*}{119.9} & \multirow {2}{*}{63.4} & \multirow {2}{*}{0.00} & \multirow {2}{*}{5.40}\\ 
			&  & \% stop & 40.5&41.0&40.2&39.3&40.2&39.9 & & & &\\
			& \multirow {2}{*}{Prior 3} & \% reject & 10.1&10.0&10.2&10.3&9.8&10.1 & \multirow {2}{*}{118.1} & \multirow {2}{*}{57.9} & \multirow {2}{*}{0.00} & \multirow {2}{*}{5.42}\\
			&  & \% stop & 43.5&44.1&42.9&42.3&43.4&42.5 & & & &\\
			& \multirow {2}{*}{Prior 4} & \% reject & 10.3&10.8&9.9&10.6&9.6&9.8 & \multirow {2}{*}{119.9} & \multirow {2}{*}{62.7} & \multirow {2}{*}{0.00} & \multirow {2}{*}{5.37}\\
			&  & \% stop & 40.7&40.9&40.1&39.2&40.7&39.8 & & & &
			\\\\  
            2 &  & True RRs & \textbf{0.4} & 0.2 & 0.2 & 0.2 & 0.2 & 0.2 & & & 1 & 5\\
            & \multirow {2}{*}{Prior 1} & \% reject & 81.6  &  13.4   &  13.6   &  12.9   &  13.7   &  13.9 & \multirow {2}{*}{126.7} & \multirow {2}{*}{41.9} & \multirow {2}{*}{0.82} & \multirow {2}{*}{4.33}\\ 
			&  & \% stop &  3.2  &  34.6  &  33.1  & 34.0    & 34.2  &  33.7 & & & &\\
            & \multirow {2}{*}{Prior 2} & \% reject &79.1&14.2&14.3&14.1&15.0&14.9 & \multirow {2}{*}{126.5} & \multirow {2}{*}{36.2} & \multirow {2}{*}{0.79} & \multirow {2}{*}{4.28}\\ 
			&  & \% stop &3.7&34.4&33.7&34.7&34.6&34.1& & & &\\
			& \multirow {2}{*}{Prior 3} & \% reject & 77.3&15.8&16.2&15.8&16.3&16.5 & \multirow {2}{*}{126.0} & \multirow {2}{*}{44.7} & \multirow {2}{*}{0.82} & \multirow {2}{*}{4.41}\\
			&  & \% stop & 5.1&35.4&34.4&35.6&34.7&34.9 & & & &\\
			& \multirow {2}{*}{Prior 4} & \% reject & 79.1&15.0&14.4&14.8&15.6&15.7 & \multirow {2}{*}{126.6} & \multirow {2}{*}{34.3} & \multirow {2}{*}{0.80} & \multirow {2}{*}{4.22}\\
			&  & \% stop & 3.9&34.4&33.6&34.7&33.8&33.7 & & & &
			\\\\  
            3 &  & True RRs & \textbf{0.4} & \textbf{0.4} & 0.2 & 0.2 & 0.2 & 0.2 & & & 2 & 4\\ 
		    & \multirow {2}{*}{Prior 1} & \% reject & 84.5   & 85.1  & 16.2  &  14.9  & 15.8 & 15.0 & \multirow {2}{*}{131.5} & \multirow {2}{*}{34.7} & \multirow {2}{*}{1.70} & \multirow {2}{*}{3.37}\\ 
			&  & \% stop & 2.6 & 3.0 & 29.7 & 31.0 & 28.8 & 29.9 & & & &\\
            & \multirow {2}{*}{Prior 2} & \% reject &85.1&85.7&16.9&16.7&16.5&15.9& \multirow {2}{*}{131.6} & \multirow {2}{*}{33.4} & \multirow {2}{*}{1.71} & \multirow {2}{*}{3.33}\\ 
			&  & \% stop &3.2&2.4&29.0&30.2&29.7&29.8& & & &\\
			& \multirow {2}{*}{Prior 3} & \% reject & 86.3&86.2&19.7&18.5&19.9&19.3 & \multirow {2}{*}{131.5} & \multirow {2}{*}{38.1} & \multirow {2}{*}{1.67} & \multirow {2}{*}{3.44}\\
			&  & \% stop & 3.5&3.5&29.3&30.3&29.1&29.7 & & & &\\
			& \multirow {2}{*}{Prior 4} & \% reject &  85.8&86.2&18.3&17.0&17.9&18.0& \multirow {2}{*}{131.7} & \multirow {2}{*}{33.3} & \multirow {2}{*}{1.73} & \multirow {2}{*}{3.26}\\
			&  & \% stop & 2.8&3.0&28.9&30.0&28.9&29.4 & & & &
			\\\\  
            4 & & True RRs & \textbf{0.4} & \textbf{0.4} & \textbf{0.4} & 0.2 & 0.2 & 0.2 & & & 3 & 3\\ 
            & \multirow {2}{*}{Prior 1} & \% reject & 87.6   &  87.9  &  87.4  &  17.4  &  18.0  &  17.1 & \multirow {2}{*}{135.9} & \multirow {2}{*}{37.5} & \multirow {2}{*}{2.63} & \multirow {2}{*}{2.48}\\ 
			&  & \% stop &2.1&2.1&2.2&25.9&24.6&24.5& & & &\\
            & \multirow {2}{*}{Prior 2} & \% reject &88.6&88.8&88.7&19.2&20.2&18.8 & \multirow {2}{*}{136.3} & \multirow {2}{*}{37.4} & \multirow {2}{*}{2.66} & \multirow {2}{*}{2.42}\\ 
			&  & \% stop &2.2&2.3&2.2&24.3&23.1&23.3& & & &\\
			& \multirow {2}{*}{Prior 3} & \% reject & 89.6&89.5&89.6&22.0&22.5&21.6 & \multirow {2}{*}{136.3} & \multirow {2}{*}{37.5} & \multirow {2}{*}{2.58} & \multirow {2}{*}{2.52}\\
			&  & \% stop & 2.1&2.5&2.4&24.1&22.9&22.8 & & & &\\
			& \multirow {2}{*}{Prior 4} & \% reject & 89.0&89.0&88.7&20.4&20.8&20.0 & \multirow {2}{*}{136.5} & \multirow {2}{*}{36.7} & \multirow {2}{*}{2.68} & \multirow {2}{*}{2.38}\\
			&  & \% stop & 2.1&2.3&2.2&23.4&22.2&23.2 & & & &
			\\\\  
            5 &  & True RRs & \textbf{0.4} & \textbf{0.4} & \textbf{0.4} & \textbf{0.4} & 0.2 & 0.2 & & & 4 & 2\\ 
			& \multirow {2}{*}{Prior 1} & \% reject & 88.7&89.0&89.9&88.9&19.2&18.1 & \multirow {2}{*}{139.5} & \multirow {2}{*}{41.2} & \multirow {2}{*}{3.56} & \multirow {2}{*}{1.63}\\ 
			&  & \% stop & 1.6 & 1.4 & 1.5 & 1.5 & 20.2 &20.3 & & & &\\
            & \multirow {2}{*}{Prior 2} & \% reject &90.1&90.3&91.1&90.7&20.9&21.8 & \multirow {2}{*}{139.8} & \multirow {2}{*}{41.6} & \multirow {2}{*}{3.59} & \multirow {2}{*}{1.58}\\ 
			&  & \% stop &1.6&1.5&1.7&1.6&17.8&17.6& & & &\\
			& \multirow {2}{*}{Prior 3} & \% reject & 90.0&90.5&91.4&90.9&24.6&24.3 & \multirow {2}{*}{140.0} & \multirow {2}{*}{39.0} & \multirow {2}{*}{3.49} & \multirow {2}{*}{1.65}\\ 
			&  & \% stop & 1.7&1.5&1.6&1.6&16.8&16.8 & & & &\\
			& \multirow {2}{*}{Prior 4} & \% reject &89.6&90.3&91.2&90.2&23.2&22.8 & \multirow {2}{*}{140.1} & \multirow {2}{*}{39.2} & \multirow {2}{*}{3.62} & \multirow {2}{*}{1.53}\\ 
			&  & \% stop & 1.5&1.4&1.4&1.3&16.8&16.8 & & & &
			\\\\  
			6 &  & True RRs & \textbf{0.4} & \textbf{0.4} & \textbf{0.4} & \textbf{0.4} & \textbf{0.4} & 0.2 & & & 5 & 1\\
            & \multirow {2}{*}{Prior 1} & \% reject & 90.5&91.1&90.7&90.3&89.9&25.0 & \multirow {2}{*}{142.5} & \multirow {2}{*}{43.0} & \multirow {2}{*}{4.56} & \multirow {2}{*}{0.73}\\ 
			&  & \% stop & 1.0&0.9&1.0&0.9&1.2&10.3 & & & &  \\ 
            & \multirow {2}{*}{Prior 2} & \% reject &91.1&91.6&91.2&90.8&90.6&25.6& \multirow {2}{*}{142.4} & \multirow {2}{*}{43.6} & \multirow {2}{*}{4.55} & \multirow {2}{*}{0.74}\\  
			&  & \% stop &0.8&1.1&1.1&0.9&1.2&10.7& & & &  \\ 
			& \multirow {2}{*}{Prior 3} & \% reject & 92.0&92.1&92.0&91.7&91.2&29.1 & \multirow {2}{*}{142.4} & \multirow {2}{*}{51.1} & \multirow {2}{*}{4.55} & \multirow {2}{*}{0.82}\\ 
			&  & \% stop & 1.0&1.0&1.1&0.9&1.2&10.2 & & & &\\
			& \multirow {2}{*}{Prior 4} & \% reject & 91.6&91.8&92.0&91.2&91.1&27.7 & \multirow {2}{*}{142.5} & \multirow {2}{*}{43.3} & \multirow {2}{*}{4.58} & \multirow {2}{*}{0.72}\\ 
			&  & \% stop & 1.0&0.9&1.0&0.9&1.1&10.1 & & & &			
			\\
			\hline
		\end{tabular}
\end{table*}
\FloatBarrier

\begin{table*}[htb]
\centering
\caption{
Comparison between the various approaches considered. The number of ``$\star$'' indicates the preference level (more ``$\star$'' indicates higher level of preference), and ``\checkmark'' indicates that a approach is recommended.}\label{table3}
	\begin{tabular}{cccccc}
			\hline
			Method &  Power & \makecell{Type I Error\\ Control} & \makecell{Design\\ Complexity} & Overall\\
            \hline
            Independent & $\star$ & $\star\star\star\star\star$ &  $\star\star\star$  &  \checkmark\\ 
            BHM & $\star\star\star$ & $\star$ &  $\star\star\star$  &  \\ 
            EXNEX & $\star\star\star$ & $\star\star$ &  $\star$  &  \\ 
            Liu's & $\star\star$ & $\star\star\star$ &  $\star$  &  \\ 
            CBHM & $\star\star$ & $\star\star\star\star$ &  $\star\star$  &  \checkmark\\ 
            \hline
		\end{tabular}
\end{table*}

\section{Discussion}\label{discussion}
We discuss several representative Bayesian methods for basket trials in early-phase oncology studies that borrow information across indications, including BHM, EXNEX and Liu's approach, and 
propose a novel method called CBHM, which conducts flexible information borrowing according to pairwise sample similarity of the indications. 
Simulation results show the advantage as well as the drawback of information borrowing: it potentially provides substantially higher power, but the type I error rate will be inflated as well. 
Among the considered information borrowing approaches, BHM only demonstrates clear advantage in terms of power when a large proportion (e.g. 5 out of 6) of the indications are sensitive, and has the highest inflation in type I error rate that can be unacceptable even for early-phase trials. 
Our CBHM approach has the most robust performance, with a power similar to that of EXNEX or Liu's approach depending on the data scenario, and type I error control that can be substantially better than that of BHM, EXNEX and Liu's approach.

In the early-phase basket trials, type I error control is not our primary focus, and thus various approaches have been proposed to improve the power by sacrificing strong control of type I error rate. 
However, the inflated type I error rate should still be concerned while focusing on improving the detection power. According to our simulation results, BHM and EXNEX can have type I error rates inflated from $10\%$ to $33.6\%- 63.1\%$ when 5 of the 6 indications are sensitive. This is a huge inflation that can raise serious concerns about the validity of the conclusions, suggesting that we should be cautious when determining if it is worth conducting an information borrowing approach. 
In this sense, the advantage of CBHM in terms of type I error control can be quite meaningful in clinical application.

In general, certain things should be kept in mind when implementing a Bayesian information borrowing approach. First, the prior specification needs careful consideration due to limited sample size. 
For CBHM, for example, the prior for the correlation range parameter, $\phi$, should be specified based on the null and target response rates, distance measure and correlation function, to induce a reasonable amount of borrowing. 
Second, in the current setting, we assume that all indications have the same null and target response rates for simplicity. In reality, the trial setting can be more complex. For example, the null response rates, $q_{0,i}$'s, and the target response rates, $q_{1,i}$'s, may vary by indication. 
Our CBHM approach, however, can be easily extended to apply to this general case after slight modification. 
Given these practical concerns, we provide CBHM with a detailed instruction for the prior specification and trial process in a general setting. This type of general guidance is not provided by most existing approaches, including EXNEX and Liu's approach, which ensures the applicability of CBHM in a wider range of trials with more complex settings. In summary, for a specific trial, simulation studies should be conducted to determine the prior for the model parameters and the optimal approach for analysis (e.g. independent analysis or information borrowing approaches such as BHM or CBHM), in order to have our desired power and type I error rate.

We conducted CBHM using the R package ``rjags'', which on average takes approximately 17 minutes to complete the simulations under one data scenario if conducting parallel computing across simulations using 24 cores. Although rjags is sufficient for our current simulation, we can consider other software, such as STAN (Carpenter et al, 2017), 
for future application, which can potentially improve the computational efficiency of the modeling process. 
The current CBHM approach restrains borrowing between heterogeneous indications by assigning a correlation close to zero between indications that have large difference in sample response rates.
However, it still leads to inflated type I error rate, and, ideally, we would like to avoid information borrowing between a sensitive and an insensitive indication. 
This inspires us to modify CBHM to achieve more strict type I error control, for example, to truncate the correlation between two indications when the difference in sample response rates exceeds a threshold. 
We leave this extension of CBHM for future work.

\section{Software}
The R code for simulation studies and the implementation of CBHM in a general trial setting, as well as the simulated data sets
that were used to produce the presented simulation results, are available at 

\section*{Acknowledgements}
We thank Dr. Joseph Koopmeiners from Devision of Biostatistics at the University of Minnesota for the helpful comments and suggestions on the paper. 




\section*{Supporting information}
Additional supporting information may be found online in the Supporting Information Section at the end of the article.

 \newpage
\appendix

\section*{Appendix A. Prior specification for BHM}
For the conventional BHM, it is assumed that $Y_{ij} \stackrel{iid}{\sim} \text{Ber}(p_i)$, and the log odds parameters, $\{\theta_i = \log(p_i/(1-p_i))$, $i=1,2,\ldots,I\}$, follow a common prior $N(\theta_0,\sigma^2)$, where $\theta_0\sim N(g_0,\sigma^2_0)$ with $g_0=\log(q_0/(1-q_0))$, $\sigma^2_0=1000$, and $\sigma^2\sim \text{IG}(a_{\sigma^2},b_{\sigma^2})$ with $a_{\sigma^2}=b_{\sigma^2}=0.001$. Here $\text{IG}(a,b)$ denotes the inverse Gamma distribution with shape parameter $a$ and rate parameter $b$. 

\section*{Appendix B. Prior specification for EXNEX}
Recall that EXNEX has the following model setup:
\begin{align*}
Y_{ij} \sim & \text{Ber}(p_i),\\
\theta_i =  \log & \frac{p_i}{1-p_i},\\
\theta_i = \delta_i M_{i1} & + (1-\delta_i) M_{i2},\nonumber\\
\delta_i \sim \text{Ber}(\pi_i), \text{  }
M_{i1} \sim N & (\mu_0,\sigma^2_0), \text{  } M_{i2} \sim N(\mu_i,\sigma^2_i).
\end{align*} 
The prior specification for EXNEX used in this paper follows the one that was used in the original paper (Neuenschwander et al., 2016):
\begin{align*}
\mu_0 \sim & N(0,\tau^2_{0}),\nonumber\\
\sigma^2_0 \sim TrN(0,& 100,(0.001,+\infty)), \nonumber\\
(\pi_i,1-\pi_i)  & \sim \text{Dir}(\lambda_1,\lambda_2),
\end{align*}
where $TrN(a,b,(L,U))$ denotes the truncated normal distribution on the interval $(L,U)$, and \\ $\text{Dir}(\lambda_1,\lambda_2,\ldots,\lambda_K)$ denotes the Dirichlet distribution with probability parameters $(\lambda_1,\lambda_2,\ldots,\lambda_K)$. 
Note that $\pi_i$'s and $1-\pi_i$'s, i.e. the weights of EX and NEX components, respectively, can be assumed as fixed and pre-specified by users, or alternatively, as done in our simulation studies, can be inferred along with the other parameters 
via a Dirichlet prior. 
The specification for hyperparameters follows Neuenschwander et al. (2016): $\tau^2_{0}=5$, 
$\mu_i=\log(q_0/(1-q_0))$, and $\sigma^2_i=1/0.15$, $i=1,2,\ldots,I$. 
The current EXNEX model assumes only one EX component, which, in some cases, may be too restrictive. The extension to more than one EX component is technically straightforward, but will require careful consideration for the choice of mixture weights and priors (details can be found in Neuenschwander et al., 2016). 
Under the simulation setting presented in this paper, a single EX component is adequate to illustrate the performance of EXNEX.

\section*{Appendix C. Prior specification for Liu's two-path design}
For Liu's two-path design, we set the threshold for the test statistics of the homogeneous test as $\gamma=0.2$, which was used in the original paper (Liu et al., 2017). 
The heterogeneous path follows Simon's two-stage design, and there is no model parameter to estimate or calibrate.
For the homogeneous path, we set the probability threshold for futility stopping at interm analysis to $C=0.5$, which was also used in Liu et al. (2017). 
The model and prior specification for stage one is described in Section 2.2.2 of the main manuscript. For stage two, a BHMM with two mixture components is applied to the final data:
\begin{align*}
Y_{ij} \sim & \text{Ber}(p_i),\nonumber\\
\theta_i = &\log\frac{p_i}{1-p_i} \\
\theta_i = \delta_i M_{i1} & + (1-\delta_i) M_{i2},\nonumber\\
\delta_i\sim \text{Ber}(\pi_i),\text{  }
M_{i1} \sim N & (\mu_1, \sigma^2_1), \text{  } M_{i2} \sim N(\mu_2,\sigma^2_2),
\end{align*}
and the prior specification follows Liu et al. (2017):
\begin{align*}
\mu_1 \sim N(g_1,\tau^2_{1}),\text{ } & \mu_2 \sim N(g_2,\tau^2_{2}),\nonumber\\
\sigma^2_1 \sim \text{IG}(a_1,b_1), \text{ } & \sigma^2_2 \sim \text{IG}(a_2,b_2).
\end{align*}
In the simulation studies in Liu et al. (2017), $g_1$ and $g_2$ were specified to be $-2.2$ and $-1.1$, respectively, which, in their simulation setting, are $g_1=\log(q_0/(1-q_0))$, and $g_2=\log(q_1/(1-q_1))$, respectively. 
In our simulation setting, we used the same strategy for prior specification and set $g_1=\log(q_0/(1-q_0))=-1.38$, and $g_2=\log(q_1/(1-q_1))=-0.41$. For the rest of the hyperparameters, we set $\tau^2_1=1/0.42$, $\tau^2_2=1/0.57$, $a_1=a_2=b_1=b_2=0.1$, and $\pi_i=0.5$, $i=1,2,\ldots,I$, which were also used in Liu et al. (2017).

\section*{Appendix D. Detailed discussion on CBHM}
\subsection*{D.1. Prior specification for CBHM in the presented simulation studies}
Recall that the model setup for CBHM is:
\begin{align*}
Y_{ij} \sim & \text{Ber}(p_i),\nonumber\\
\theta_i = & \log\frac{p_i}{1-p_i}, \\
\theta_i =\theta_0 & + \eta_i + \epsilon_i, \nonumber\\
\bm{\eta} \sim \mathcal{MVN}(\bm{0},\sigma^2\bm{R}(\phi)) &, \text{  }
\epsilon_i \stackrel{i.i.d.}{\sim} N(0,\tau^2).\label{cbhm}
\end{align*}

Now we introduce how we specified the prior for CBHM used in the simulation studies discussed in Section 3.2 of the main manuscript.
Briefly, we first specify the prior for the correlation range parameter $\phi$ based on the distance measure, a distance threshold $d^t$, a correlation threshold $\varrho$, and the choice for correlation function. 
We then specify the prior for the rest of the parameters.

For the prior specification of $\phi$, the first step is to select a distance measure. 
Theoretically, any well-defined distance measure for probability distributions can be used.

We first use B distance as an example, which, in our beta-binomial case, can be written as:
\begin{align*}
d^{B}_{i,j}=-\log\frac{B\left(\frac{r_i+r_j+2}{2},\frac{n_i+n_j-r_i-r_j+2}{2}\right)}{\sqrt[]{B(1+r_i,1+n_i-r_i)B(1+r_j,1+n_j-r_j)}},
\end{align*}
where $B(a,b)$ denotes the Beta function, and the other notations are introduced in Section 2.3.1 of the main manuscript. 

Figure S1 shows the sample distribution of B distance between two indications in the following scenarios: (1) $p_1=p_2=q_0$, (2) $p_1=p_2=q_1$, (3) $p_1=q_0$, $p_2=q_1$, where $q_0$ and $q_1$ denote the null response rate and target response rate, respectively. 
\begin{figure*}[htbp!]
\centering\includegraphics[width=430pt]{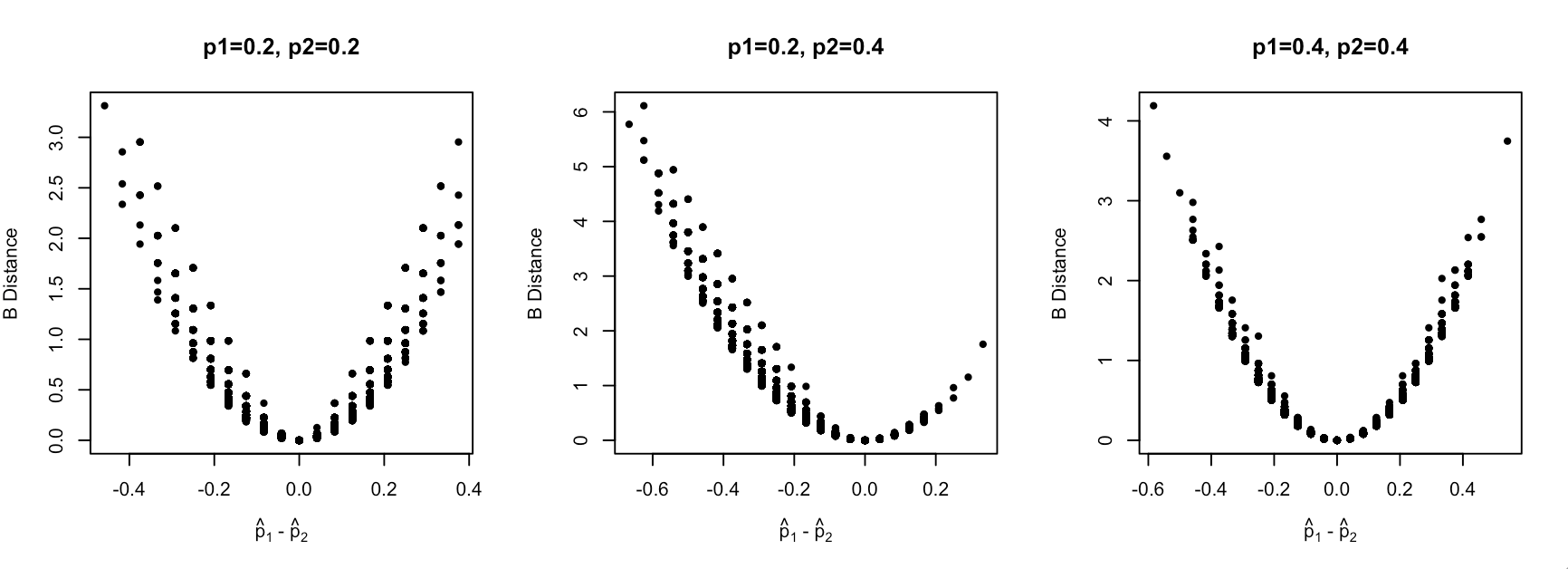}
    \caption{ Example plots showing the sample distribution of B distance between two indications when (1) $p_1=p_2=q_0$, (2) $p_1=p_2=q_1$, and (3) $p_1=q_0$, $p_2=q_1$, respectively. Results were obtained from 10000 simulations in each case. Sample sizes are fixed to $n_1=n_2=24$, the null response rate is set to $q_0=0.2$ and the target response rate is set to $q_1=0.4$. 
    }
\label{fig1_supp}
\end{figure*}
In the two homogeneous cases (1) and (3) (sub-figures 1 and 3 of Figure S1), the $95\%$ quantile of the sample distance is $d^t=0.995$, which will be used as a threshold for distance that helps us determine if a pair of indications is homogeneous: if the distance is smaller than $d^t$, then we say that the two indications are homogeneous and that we want to conduct strong information borrowing between them by assigning them with a high prior correlation. 
We also set a correlation threshold, $\varrho$, for information borrowing: we would like to assign a prior correlation higher than $\varrho$ between homogeneous indications, and a prior correlation lower than $\varrho$ between heterogeneous indications. Simulation results show that a value between 0.3 and 0.5 can be a reasonable choice for $\varrho$.

Finally, we choose the correlation function. Under exponential correlation, given the distance threshold $d^t=0.995$, a correlation within the interval $[0.3,0.5]$ corresponds to a $\phi$ within the interval $[-\log(0.5)/d^t,-\log(0.3)/d^t]=[0.70,1.21]$, meaning that any value within this interval can be a choice for the prior mean of $\phi$. In the simulation studies presented in Section 3.2 of the main manuscript, we set this prior mean to 1. 
Note that for B distance, the squared correlation function is unrealistic because of its strong smoothness. A simple example to illustrate this point is that if we use the squared exponential correlation with $\phi=1$, 
when $p_1=p_2=0.2$, approximately $33.54\%$ of the sample distance will fall into $[0,0.032]$, meaning that the prior correlations we assign to $33.54\%$ of the pairs of indications are greater than $e^{-0.032^2\varrho} > 0.9994$. This correlation structure is unreasonable, and can easily cause computational issue for the MCMC algorithm.

We now discuss the prior specification for $\phi$. We consider a Gamma prior, $\text{G}(a,1)$, with rate parameter equal to 1, which converges to the normal distribution $N(a,\sqrt{a})$. The rate parameter is set to 1 because it only affects the scale of the distribution: if $x\sim \text{G}(a,1)$, then $\beta x \sim \text{G}(a,1/\beta)$. As discussed in the previous paragraph, we would like to set the prior mean of $\phi$ to 1, which, for the $\text{G}(a,1)$ prior, means that $a=1$, i.e. a $\text{G}(1,1)$ prior for $\phi$.

Finally, we specify the priors for the rest of the model parameters:
\begin{align*}
\theta_0 \sim N(\mu_0,\sigma^2_0), 
\sigma^2 \sim \text{IG}(c_{\sigma^2},d_{\sigma^2}),
\tau^2 \sim \text{IG} (c_{\tau^2},d_{\tau^2}),
\sigma_0^2 \sim \text{IG}(c_{\sigma_0^2},d_{\sigma_0^2}),
\end{align*}
where the hyperparameters are set to: $\mu_0=\log(((q_0+q_1)/2)/(1-(q_0+q_1)/2))$, 
$c_{\tau^2}=d_{\tau^2}=c_{\sigma^2}=d_{\sigma^2}=0.01$, and $c_{\sigma^2_0}=d_{\sigma^2_0}=0.1$, to ensure that the priors are only weakly informative. 

In the previous paragraphs, we discussed the prior specification of CBHM using the B distance. We now discuss the prior specification of CBHM using the H distance. 
Note that given a non-informative $\text{Beta}(1,1)$ prior, $p_i$ and $p_j$ have posteriors $\text{Beta}(1+r_i,1+n_i-r_i)$ and $\text{Beta}(1+r_j,1+n_j-r_j)$, respectively, and the H distance, $d^{H}_{i,j}$, can be written as:
\begin{align*}
d^{H}_{i,j}=\sqrt[]{1-\frac{B\left(\frac{r_i+r_j+2}{2},\frac{n_i+n_j-r_i-r_j+2}{2}\right)}{\sqrt[]{B(1+r_i,1+n_i-r_i)B(1+r_j,1+n_j-r_j)}}}.
\end{align*}

Following a similar procedure as the one used for B distance, we first simulate sample distribution of H distance in the following scenarios: (1) $p_1=p_2=q_0$, (2) $p_1=p_2=q_1$, (3) $p_1=q_0$, $p_2=q_1$ (see Figure S2).
\begin{figure*}[htb]
\centering\includegraphics[width=430pt]{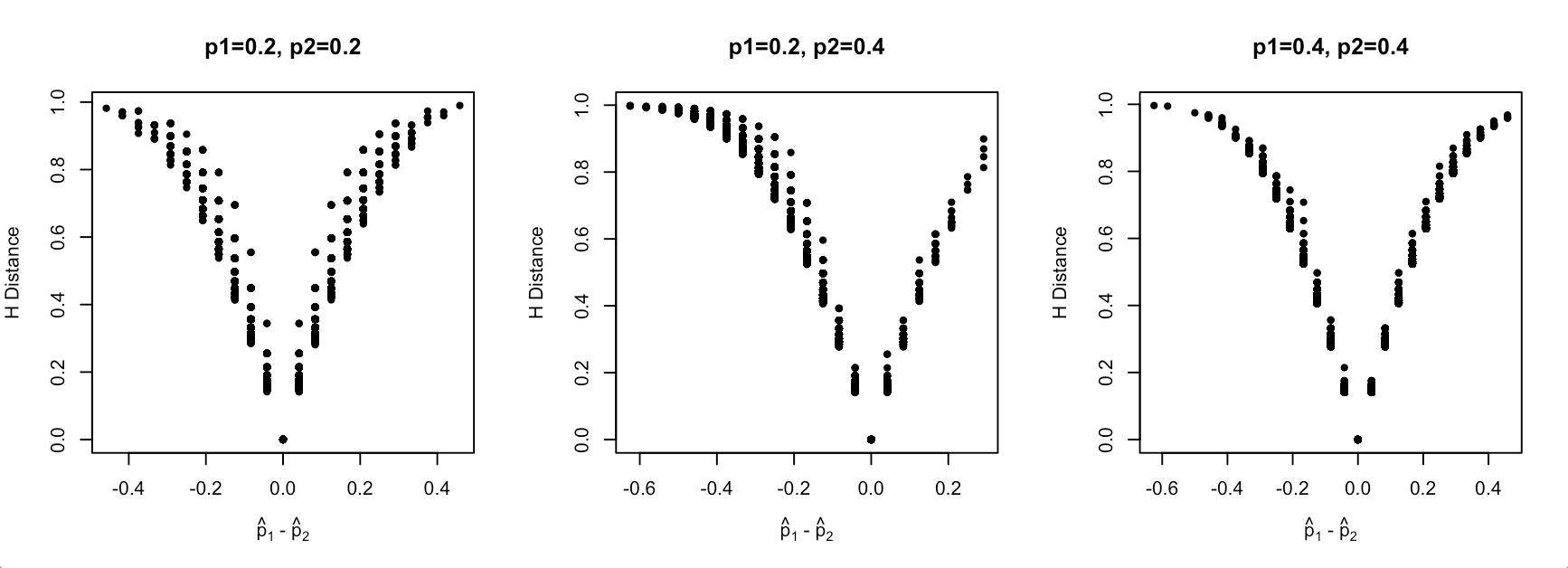}
    \caption{Example plots showing the sample distribution of H distance between two indications when $p_1=p_2=q_0$, $p_1=p_2=q_1$, and $p_1=q_0$, $p_2=q_1$, respectively. Results were obtained from 10000 simulations in each case. Sample sizes are fixed to $n_1=n_2=24$, the null response rate is set to $q_0=0.2$ and the target response rate is set to $q_1=0.4$.}
\label{fig2_supp}
\end{figure*}
In the two homogeneous cases (1) and (3) (sub-figures 1 and 3 of Figure S2), the $95\%$ quantile of the sample distance is $d^t=0.793$, which will be used as the distance threshold. 
For H distance, we also use the exponential correlation function since the squared correlation is still not appropriate due to the reason similar to the one when using B distance. We still set the correlation threshold to  $\varrho\in [0.3,0.5]$. Given distance $d^t=0.793$, the prior mean of $\phi$ can be selected within the interval $[-\log(0.5)/d^t,-\log(0.3)/d^t]=[0.87,1.52]$. In the simulation studies discussed in this paper, we set the prior mean of $\phi$ to 1.5, which, for the $\text{G}(a,1)$ prior, means that $a=1.5$. The priors for the rest of the model parameters are specified as:
\begin{align*}
\theta_0 \sim N(\mu_0,\sigma^2_0),
\sigma^2 \sim \text{IG}(c_{\sigma^2},d_{\sigma^2}),
\tau^2 \sim \text{IG} (c_{\tau^2},d_{\tau^2}),
\sigma_0^2 \sim \text{IG}(c_{\sigma_0^2},d_{\sigma_0^2}),
\end{align*}
where the hyperparameters are set to: $\mu_0=\log(((q_0+q_1)/2)/(1-(q_0+q_1)/2))$, 
$c_{\tau^2}=d_{\tau^2}=c_{\sigma^2}=d_{\sigma^2}=0.01$, and $c_{\sigma^2_0}=d_{\sigma^2_0}=0.1$.

\begin{figure*}[htb]
\centering\includegraphics[width=430pt]{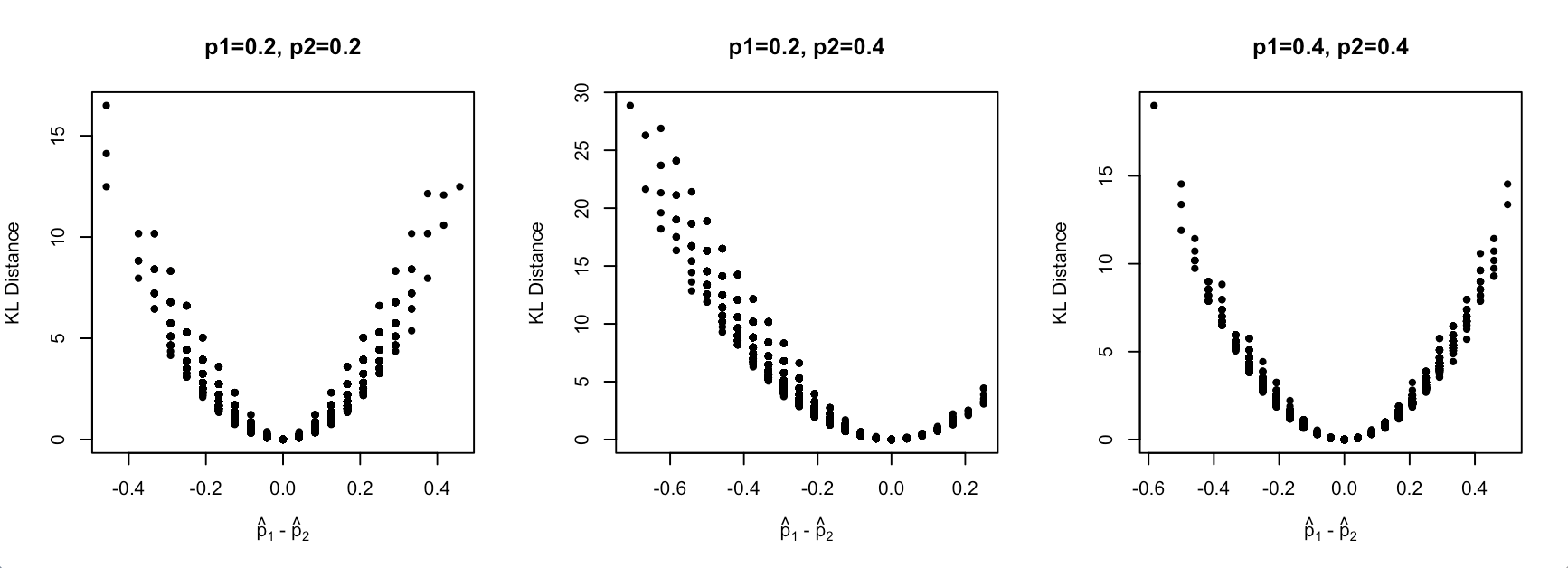}
    \caption{Example plots showing the sample distribution of KL distance between two indications when $p_1=p_2=q_0$, $p_1=p_2=q_1$, and $p_1=q_0$, $p_2=q_1$, respectively. Results were obtained from 10000 simulations in each case. Sample sizes are fixed to $n_1=n_2=24$, the null response rate is set to $q_0=0.2$ and the target response rate is set to $q_1=0.4$.}
\label{fig3_supp}
\end{figure*}
We now discuss the prior specification of CBHM using the KL distance. The simulated sample distribution in the various data scenarios is shown in Figure S3. 
In the two homogeneous cases (sub-figures 1 and 3 of Figure S3), the $95\%$ quantile of the sample distance is $d^t=2.710$, which will be used as the distance threshold. Since KL distance changes much faster than B distance and H distance as $\hat{p_1}-\hat{p_2}$ changes, we consider a more smooth correlation function, the squared exponential correlation function. Given $d^t=2.710$, the prior mean of $\phi$ can be selected within the interval $[\sqrt{-\log(0.5)/d^t},\sqrt{-\log(0.3)/d^t}]=[0.094,0.164]$. Theoretically, we could still use a $\text{G}(a,1)$ prior for $\phi$ with $a\in [0.094,0.164]$.
However, if using the R package ``rjags'' to run MCMC, some adjustments should be made to solve the potential issue that in some rare cases, the draw of $\phi$, $\sigma^2$ and $\tau^2$ leads to singular between-indication covariance matrix, $\sigma^2\bm{R}(\phi)$. This is because in some MCMC iterations the sampled $\phi$ or $\sigma^2$ is too close to 0. 
We can solve this problem by writing our own MCMC algorithm instead of using rjags. For easy implementation using rjags, we 
modify the prior distribution and constrain the scale of the hyperparameters to avoid having MCMC draws that may generate singular $\sigma^2\bm{R}(\phi)$: 
instead of using the $\text{G}(a,1)$ prior, we use a  $\text{Unif}(a,b)$ prior for $\phi$ with $a>0$. In the two homogeneous cases (sub-figures 1 and 3 of Figure S3), the minimum value of the nonzero distances is $d^t=0.0727$. To ensure that the corresponding correlation is less than 0.999, $\phi$ has to be larger than 0.189. We therefore set a $\text{Unif}(0.189,0.5)$ prior for $\phi$. 
The prior for the rest of the model parameters is specified as follows:
\begin{align*}
\theta_0 \sim N(\mu_0,\sigma^2_0) &, \phi \sim \text{Unif}(a,b),\\
\sigma^2 \sim \text{IG}(c_{\sigma^2},d_{\sigma^2}),
\tau^2 \sim \text{IG} & (c_{\tau^2},d_{\tau^2}),
\sigma_0^2 \sim \text{IG}(c_{\sigma_0^2},d_{\sigma_0^2}),
\end{align*}
where the hyperparameters are set to: $\mu_0=\log(((q_0+q_1)/2)/(1-(q_0+q_1)/2))$, 
$a=0.189$, $b=0.5$,
$c_{\tau^2}=c_{\sigma^2}=2$, $d_{\sigma^2}=3$, $d_{\tau^2}=4$, and $c_{\sigma^2_0}=d_{\sigma^2_0}=0.1$.
Note that the prior for $\tau^2$ and $\sigma^2$ should also be carefully chosen to ensure than no extremely small valued-draws are sampled in MCMC. 
Here we use a $\text{Unif}  (2,4)$ prior for $\tau^2$ and a $\text{Unif}(2,3)$ prior for $\sigma^2$.

Overall, the prior specification given squared correlation function is more complex and harder than that given exponential correlation function. We have discussed the prior specification given squared correlation function so as to show that CBHM is applicable using various types of correlation functions, and that the performance of CBHM is robust with respect to the choice of distance measure and correlation function if the prior is chosen carefully.
For the real-world application of CBHM, however, we recommend to use the B distance function along with exponential correlation function given the easy implementation.

\subsection*{D.2. An instruction for the implementation of CBHM in a general trial setting}

\subsubsection*{D.2.1 Data preprocessing}

Before applying the CBHM model, some adjustments to the data are potentially needed to ensure that the between-indication distance matrix is invertible. For example, if there are two indications that have equal sample response rates, the induced distance matrix will be singular. To solve this problem, we add a small value, $\varepsilon$, to the number of responders for one indication. 
In a more general case where $k\geqslant 2$ indications have equal sample response rates, we redefine the corresponding $r_i$s as $\{r_i+\epsilon i,i=1,2,\ldots,k\}$ to make the response rates distinguishable from each other. Note that the value of $\varepsilon$ can be adjusted according to the total number of indications, $I$, and the difference between the null response rate, $q_0$, and the target response rate, $q_1$. In our simulation studies, we set $\varepsilon = 3(q_1-q_0)/I$.

\subsubsection*{D.2.2 Prior specification}
Previously in Section D.1, we discussed how to specify the prior for CBHM assuming different distance measures and correlation functions.  
In the following instruction for the prior specification of CBHM, we propose to use the B distance along with the exponential correlation function because of the promising model performance it leads to and simple procedure required for prior specification.

We first discuss prior specification for the correlation range parameter $\phi$. Following a procedure similar to the one described in Section D.1, we consider a $\text{G}(a,1)$ prior for $\phi$, and we propose the following algorithm to determine the value for $a$, with R code available at 

\begin{algorithm}
\caption{Determining the value of $a$ in the $\text{G}(a,1)$ prior for $\phi$}\label{algorithm}
\KwIn{Input: $M$, $N$, $q_0$, $q_1$, $\alpha$, $I$, $\varrho_{lb}$, $\varrho_{ub}$}
\KwOut{Output: $a$}
Determine the distance threshold $d^t$:\\
(I) for $1\leq i<j\leq I$:\\
under each of the two homogeneous scenarios: (1) $p_i=q_{0}[i]$, $p_j=q_{0}[j]$, and (2) $p_i=q_{1}[i]$, $p_j=q_{1}[j]$, simulate M sample distances between the $i^{th}$ and the $j^{th}$ indications, and denote the set of sample distances simulated in the two scenarios as $\{d_{i,j}^{1,1},\ldots,d_{i,j}^{1,M},\ldots,d_{i,j}^{2,1},\ldots,d_{i,j}^{2,M}\}$;
\linebreak
(II) calculate the $1-\alpha$ quantile of $\{d_{i,j}^{k,l},1\leq i<j\leq I,k=1,2,l=1,2,\ldots,M\}$, which is denoted as $d^t$.

Determine the lower and upper bound of $a$:\linebreak
$a_{lb}=-\log(\varrho_{ub})/d^t$,\linebreak
$a_{ub}=-\log(\varrho_{lb})/d^t$.

Generate $a$ from $\text{Unif}(a_{lb},a_{ub})$.
\end{algorithm}
\FloatBarrier
Here $M$ denotes the number of simulations in each of the two homogeneous scenarios, $N$ denotes the vector of the number of patients enrolled in each indication group, 
$q_0$ and $q_1$ denote the vector of the null response rates and target response rates, respectively, for the $I$ indications in the trial, 
$\varrho_{lb}$ and $\varrho_{ub}$ denote the lower bound and upper bound, respectively, for the correlation threshold $\varrho$. A recommended set of input is: $M=5000$, $\alpha = 0.05$, $\varrho_{lb}=0.3$, and $\varrho_{ub}=0.5$.

We recommend the following set of priors for $\theta_0$, $\sigma^2$, $\tau^2$ and $\sigma^2_0$:
\begin{align*}
\theta_0 \sim N(\mu_0,\sigma^2_0), 
\sigma^2 \sim \text{IG}(c_{\sigma^2},d_{\sigma^2}),
\tau^2 \sim \text{IG} (c_{\tau^2},d_{\tau^2}),
\sigma_0^2 \sim \text{IG}(c_{\sigma_0^2},d_{\sigma_0^2}),
\end{align*}
where the hyperparameters are set to:
\begin{align*}
\mu_0 & =\text{log} \frac{\frac{\sum_{i=1}^K q_{0,i}+\sum_{i=1}^K q_{1,i}}{2K}}
{1-\frac{\sum_{i=1}^K q_0+\sum_{i=1}^K q_1}{2K}}, \\ c_{\tau^2}=d_{\tau^2} & = c_{\sigma^2}=d_{\sigma^2}=0.01,  c_{\sigma^2_0}=d_{\sigma^2_0}=0.1.
\end{align*}

\subsubsection*{Step 3: Trial process}
In a general trial setting, we propose a modified version of the proposed two-stage design in Section 2.3.3 of the main manuscript, where we simply change $q_{0}$ and $q_{1}$ used in the interim analysis and final decision to $q_{0,i}$'s and $q_{1,i}$'s, respectively. 
The trial design is as follows:
\begin{itemize}
\item[] 
\textit{\textbf{Step 1:}} at stage one, enroll $n^1_{i}$ patients for the $i^{th}$ indication group and collect data $D_1=\{n^1_{i},Y^1_{ij},i=1,\ldots,I,j=1,\ldots,n_i^1\}$, where $n_i^1$ denotes the sample size of the $i^{th}$ indication group in stage one.
\item[] 
\textit{\textbf{Step 2:}} at the end of stage one, apply the CBHM model to data $D_1$ and conduct interim analysis for each indication $i$: if $Pr(p_i>(q_{0,i}+q_{1,i})/2|D_1)<Q_f$, then stop the enrollment early and conclude that the $i^{th}$ indication is not sensitive to the treatment; otherwise, continue to stage two until a total of $n_i$ patients are enrolled during the trial. We denote the final data as $D=\{n_{i},Y_{ij},i:n^2_{i}>0,j=1,\ldots,n_i\}$. 
\item[] 
\textit{\textbf{Step 3:}} at the end of the trial, assess treatment efficacy on the indications that proceeded to stage two. Take the $i^{th}$ indication group as an example: if $Pr(p_i>q_{0,i}|D)>Q$, then conclude that the $i^{th}$ indication is sensitive to the treatment; otherwise, conclude that the $i^{th}$ indication is insensitive. 
\end{itemize}


\clearpage

\end{document}